\documentclass[pra,superscriptaddress,showpacs,preprint,aps]{revtex4-1}

\usepackage{epsfig}             
\usepackage{epstopdf}           
\usepackage{hyperref}           
\usepackage{color}
\usepackage{colortbl}
\usepackage{graphicx}
\usepackage{amssymb,amsmath}
\usepackage{subfigure}
\usepackage{caption}
\usepackage{enumerate}
\usepackage{gensymb}
\begin{document}

\title[ATI]{Above-threshold ionization and photoelectron spectra in atomic systems
driven by strong laser fields}


\author{Noslen Su\'arez}
\email[]{noslen.suarez@icfo.es}
\affiliation{ICFO - Institut de Ci\`encies Fot\`oniques, Av. C.F. Gauss 3, 08860 Castelldefels (Barcelona), Spain}

\author{Alexis Chac\'on}
\affiliation{ICFO - Institut de Ci\`encies Fot\`oniques, Av. C.F. Gauss 3, 08860 Castelldefels (Barcelona), Spain}

\author{Marcelo F. Ciappina}
\affiliation{Max-Planck-Institut f\"ur Quantenoptik, Hans-Kopfermann-Str. 1, 85748 Garching, Germany}

\author{Jens Biegert}
\affiliation{ICFO - Institut de Ci\`encies Fot\`oniques, Av. C.F. Gauss 3, 08860 Castelldefels (Barcelona), Spain}
\affiliation{ICREA - Instituci\'{o} Catalana de Recerca i Estudis Avan\c{c}ats, Lluis Companys 23, 08010 Barcelona, Spain}

\author{Maciej Lewenstein}
\affiliation{ICFO - Institut de Ci\`encies Fot\`oniques, Av. C.F. Gauss 3, 08860 Castelldefels (Barcelona), Spain}
\affiliation{ICREA - Instituci\'{o} Catalana de Recerca i Estudis Avan\c{c}ats, Lluis Companys 23, 08010 Barcelona, Spain}

\date{\today}

\begin{abstract}
Above-threshold ionization (ATI) results from strong field
laser-matter interaction and it is one of the fundamental
processes that may be used to extract electron structural and
dynamical information about the atomic or molecular target.
Moreover, it can also be used  to characterize the laser field
itself. Here, we develop an analytical description of ATI, which
extends the theoretical Strong Field Approximation (SFA), for both
the  direct and re-scattering transition amplitudes in atoms. From
a non-local, but separable potential, the {\em bound-free dipole} and
the {\em re-scattering transition} matrix elements are analytically
computed. In comparison with the standard approaches to the ATI
process, our {\em analytical derivation} of the re-scattering matrix
elements allows us to study directly how the re-scattering process
depends on the atomic target  and laser pulse features -- we can
turn on and off contributions having different {\em physical origins} or
corresponding to different {\em physical mechanisms}.  We compare SFA
results with the full numerical solutions of the time-dependent
Schr\"odinger equation (TDSE) within the few-cycle pulse regime. 
Good agreement between our SFA and TDSE model is found for the ATI spectrum. 
Our model captures also the strong  dependence
of the  photoelectron spectra  on the carrier  envelope phase of
the laser field.

\end{abstract}

\maketitle

\section{Introduction}

During the last three decades, advances in laser technology and
the understanding of nonlinear processes in laser-matter
interactions have led to production of few-cycle femtosecond (1 fs
$\:=10^{-15}$ s) laser pulses in the visible and  mid-infrared
regimes~\cite{MNisoli, MHemmer}. By focusing such ultrashort laser
pulses on a gas target, the atoms are subjected  to an
ultra-intense electric field, with  peak field strengths
approaching  the  binding  field inside the  atoms  themselves.
Such fields are commonly used as  a  tool  to  explore  the
interaction between  strong  electromagnetic coherent  radiation
and  an  atomic  or molecular  system  with  unprecedented spatial
and temporal resolution~\cite{Krausz:2009zz}. Phenomena such as
high-order harmonic generation
(HHG)~\cite{ALHullier1993,McPherson}, above-threshold
ionization~\cite{PRLAgostini1979}, multi-photon ionization and
multi-electron effects~\cite{RPPMainfray1991,NatKrausz2007}, are
routinely studied. These effects can be used to generate
attosecond  pulses in the extreme ultraviolet~\cite{MDrescher,
ZChang}  or even soft X-ray regime~\cite{FSilva}. They can also be
used to extract  either information  about  the laser pulse
electric field itself~\cite{NatPaulus2001}, or about the structure
of the target atom or molecule~\cite{Blaga2012,Pullen2015}.

Since electronic motion is governed by the waveform of the laser
electric field, an important quantity to describe the electric
field shape is the so-called absolute phase or
carrier-to-envelope-phase (CEP). Control over the CEP is paramount
for extracting information  about  electron dynamics, and to
retrieve structural information from atoms and
molecules~\cite{NatBaltuska2003,NatItatani2004,PRAJens2014}. For
instance, in HHG an electron is liberated from an atom or molecule
through ionization, which occurs close to the maximum of the
electric field. Within the oscillating field, the electron can
thus accelerate along oscillating trajectories, which may result
in re-collision with the parent ion, roughly when the laser field
approaches a zero value. Control over the CEP is particularly
important for HHG, when targets are driven by laser pulses
comprising only one or two optical cycles. In such situation CEP
determines the relevant electron trajectories, i.e.~the CEP
determines whether emission results in a single or in multiple
attosecond bursts of
radiation~\cite{NatBaltuska2003,NatPSola2006}.

The influence of the CEP on electron emission was also
demonstrated in an anti-correlation experiment, in which the
number of ATI electrons emitted in opposite directions was
measured~\cite{NatPaulus2001,Milosevic2006}. Since the first proof
of principle experiment~\cite{NatPaulus2001}, the stereo ATI
technique has established itself as a direct measure of the CEP,
and demonstrated its ability for single shot measurements even at
multiple kHz laser repetition rates. The sensitivity to the CEP
arises from contributions of both, bound-free and the
re-scattering  continuum-continuum transitions  of the  atomic or
molecular target, which are embedded in the photoelectron
distribution of ATI~\cite{NatItatani2004}. Hence, this mechanism
can be used to extract structural information about the target
atom or molecule.

Laser induced electron diffraction (LIED) was suggested early on
as a technique that uses the doubly differential elastic
scattering cross section to extract structural information
~\cite{Zuo1996, Lein2002, Lin2010}. Meeting the requirements to
extract structural information has, however, proven difficult due
to the stringent prerequisites on the laser parameters. During
recent years, the development of new laser sources has
dramatically advanced, leading to first
demonstrations~\cite{Xu2014,Meckel, Morishita, Xu2012,
Pullen2015}, and the successful retrieval of the bond distances in
simple diatomic molecules with fixed-angle broadband electron
scattering~\cite{Xu2014}. Recently, Pullen et
al.~\cite{Pullen2015} have exploited the full double differential
cross section to image the entire structure of a polyatomic
molecule for the first time. An important next step to exploit the
full potential of the re-collision physics is the exploitation of
the intrinsic time resolution of LIED to extract dynamic
structural information. The key for such a goal is, however, a
comprehensive and complete understanding of the ATI process and
its theoretical
description~\cite{Milosevic2006,Faisal1973,Reiss1980,Lewenstein1995,Starace2014,Becker2002,Morishita2010}.

The aim of our paper is to revisit the strong field approximation
model of M. Lewenstein for ATI for few-cycle infrared (IR) laser
pulses and to compare it with the numerical solution of the TDSE
in one (1D) and two (2D) spatial  dimensions for an atomic
system~\cite{Lewenstein1995}. For simplicity, our analytical
atomic model is based on a non-local potential, which can be
considered a short-range (SR) potential.  In order to verify the
validity of our analytical SR model, and  to understand how its
predictions compare with a true  Coulomb  potential, we
numerically  integrate the TDSE for the hydrogen atom and compute
the photoelectron energy and momentum distributions.

This article is organized as follows. In Section II, we review the
theory which describes the ATI process within the Strong Field
Approximation; in particular, we present the derivation of the
transition amplitude for both the direct and re-scattered
electrons. We develop in detail the mathematical foundations
towards the final results by starting from the Hamiltonian, which
describes the atomic system and the TDSE associated to it. In
Section III, we introduce the model for our atomic system, that
uses a particular form of a non-local short-range potential.   The
matrix elements to describe the ionization and re-scattering
processes are then  computed.  In section IV, the ATI spectra for
the 1D line and 2D case are numerically calculated  and compared
with numerical  results  obtained from TDSE calculations.  In
addition, we discuss the effect of the CEP on the spectra,
calculated from our analytical  SFA model.  Finally, in section V,
we summarize the main ideas and present our conclusions. We give
an outlook on extending this analytical model to more complex
atomic and molecular systems.

\section{Strong field approximation: transition probability amplitudes}\label{cap:Background}

The interaction of a strong electric field with an atomic or
molecular system is described within Quantum Mechanics by the time
dependent Schr\"odinger equation that  captures  both the
evolution of the (electronic) wave function and the time evolution
of the physical observables.  The numerical solution of the TDSE
offers a full quantum mechanical description  of the laser-matter
interaction processes, and has been extensively used to study
several phenomena, such as
HHG~\cite{KulanderGardee1998,TateDiMauro2007, Marcelo2012} and
ATI~\cite{Kulander1987,Muller1999,DBauer2005,Blaga2009,
RufKeitel2009} in atomic and molecular systems. However, the  full
numerical  integration of the  TDSE  in all the  degrees of
freedom of the system  is often a laborious, and sometimes an
impossible task to perform from a numerical and computational
points of view. Moreover, a physical interpretation of the
numerical TDSE results and the extraction  of information  from
the time evolved wave function is highly nontrivial for an {\it ab
initio} technique.

Hence, from a purely theoretical point of view, it would be
desirable to solve the TDSE analytically for the ionization
process. This is one of the main steps in all laser-matter
interaction phenomena, and it represents a formidable  and
challenging assignment.  Here, we discuss an alternative method to
calculate photoelectron spectra from atomic systems  by
analytically solving the  TDSE  under  the so-called Strong Field
Approximation.  This  approach  dates back to
Keldysh~\cite{Keldysh1965}, and  has  since been employed  by many
other
authors~\cite{Ammosov1986,Faisal1973,Reiss1980,Lewenstein1986,Ehlotzky1992,Lewenstein1994,Lewenstein1995,Kuchiev1999}.
 It is worth noted that SFA provides a quantum framework and extension of the, so called, ``simple man's" or ``three step"
 or ``re-collision" model, usually attributed to P. Corkum~\cite{Corkum1993}, K. Kulander~\cite{Kulander1992a,Kulander1993}
 and H. Muller (cf.~\cite{Smirnova2014} for an extensive review; for
earlier quantum formulation of  ``Atomic Antennas" see
Ref.~\cite{Kuchiev1987}; for other pioneering contributions see
Refs.~\cite{Brunel1987,Brunel1990,Corkum1989}).

\subsection*{Ionization driven by strong fields}
Let us consider an atom under the influence of an ultra-intense
laser field. In the limit when the wavelength of the laser,
$\lambda_0$, is larger compared with the Bohr radius, $a_0$
(5.29$\times 10^{-11}$ m), the electric field of the laser beam
around the interaction region can be considered spatially
homogeneous. Consequentially, the  interacting atoms  will not
experience  the spatial  dependence  of the  laser electric  field
and, hence, only its  time-variation is taken  into account.  This
is the so-called dipole approximation.  In this approximation, the
laser electric field can be written  as:
\begin{equation}
 \textbf{E}(t) = \mathcal{E}_0\:f(t) \sin(\omega_0 \:t + \phi_0)\,{\bf e}_{z}.
 \label{Eq:Efield}
\end{equation}

The field of Eq.~(\ref{Eq:Efield}) has a carrier frequency
$\omega_0 = \frac{2\pi c}{\lambda_0}$, where $c$ is the speed of
light, $\mathcal{E}_0$ the field peak amplitude or strength, and
we consider that the laser field is linearly polarized along the
$z$-direction.  ${f(t)}$ denotes the envelope of the laser pulse
and the parameter $\phi_0$, defines the CEP.

The TDSE  is defined  (atomic  units  are  used  throughout this  paper  unless otherwise stated) by:
\begin{equation}
i\frac{ \partial}{ \partial t} | \Psi(t) \rangle=\hat{H} | \Psi(t) \rangle, \label{Eq:SE}
\end{equation}
where the Hamiltonian operator, $\hat{H}$, describes the laser-atom system and is the sum of two terms, i.e.
\begin{eqnarray}
\hat{H} &= &\hat{H_0}+\hat{U}, \label{Eq:H}
\end{eqnarray}
where, $\hat{H}_0$, is the so-called laser-free Hamiltonian of the atomic or molecular system
\begin{eqnarray}
\hat{H}_0&=&- \frac{\nabla^2}{2} + \hat{V}(\textbf{r}),
\end{eqnarray}
with $\hat{V}(\textbf{r})$ the atomic or molecular potential, and
$\hat{U}=-q{\bf E}(t)\cdot {\bf r}$, is the  dipole coupling which
describes the interaction of the atomic or molecular system with
the laser radiation, written  in the length  gauge and under  the
dipole approximation.  Note that  in atomic  units,  the electron
charge, denoted  by $q$, is $q=-1$ a.u.

We shall restrict  our model to  the  low ionization  regime,
where the  SFA is valid~\cite{Keldysh1965, Faisal1973,
Reiss1980,Lewenstein1986,Lewenstein1994,Lewenstein1995} and
successfully describes the laser-matter interaction processes.
Therefore, we consider the strong field or tunneling regime, where
the Keldysh parameter  $\gamma'=\sqrt{I_p/2U_p}$ ($I_p$ denotes
the ionization potential  of the atomic or molecular system and,
$U_p=\frac{\mathcal{E}_0^2}{4\omega_0^2}$, the ponderomotive
energy taken  by the electron in the oscillating electromagnetic
field) is less than one, i.e. $\gamma'<1$. In addition,  we assume
that the  remaining  Coulomb potential, $V(\textbf{r})$,  does not
play an important role in the electron  dynamics  once the
electron  appears in the continuum.  These observations,  and the
following three statements, define the standard SFA, namely:

\begin{enumerate}[(i)]

\item The strong  field laser does not  couple with any other  bound  state.  This means that only
the ground state, $|0 \rangle$, and the continuum  states, $ |\textbf{v}\rangle$, are taken into account in the interaction process;

\item There is no depletion of the ground state,  i.e. the ponderomotive  energy is lower than the saturation
energy of the system  $(U_p < U_{sat})$; Despite this assumption,
including depletion effects, e.g. by including ionization  rates
according to the Ammosov-Delone- Krainov theory (ADK rates
\cite{Ammosov1986}) provides no a particular challenge;

\item The continuum  states  are approximated by Volkov states;  more precisely
(cf.~\cite{Lewenstein1986,Lewenstein1994,Lewenstein1995}) the
continuum-continuum matrix  elements are decomposed in the basis
of scattering  states,  corresponding  to waves with a fixed
outgoing (kinetic)  momentum  ${\bf p}_e$, into the most singular
part  and the rest, which is treated  then in a perturbative
manner~\cite{Lewenstein1995}. In such decomposition, the most
singular part  corresponds  exactly  to approximating the
scattering states  by plane waves, i.e.  Volkov solutions.
Corrections with respect to the less singular part  of the
continuum-continuum matrix  elements describe re-scattering and
re-collision events.
\end{enumerate}

Based on the statement (i), we propose a state, $|\Psi(t)
\rangle$, that describes the time-evolution of the system by a
coherent superposition of the ground, $|0 \rangle$, and the
continuum states, $
|\textbf{v}\rangle$~\cite{Lewenstein1994,Lewenstein1995}:
\begin{equation}
| \Psi(t) \rangle= e^{\textit{i}I_p\textit{t}}\bigg(a(t) |0 \rangle + \: \int{\textit{d}^3 \textbf{v} \:  \textit{b}( \textbf{v},t) |\textbf{v}\rangle} \bigg).
\label{PWavef}
\end{equation}

The factor, $a(t)$, represents  the  amplitude  of the  ground
state  which will be considered constant in time, $a(t) \approx
1$, under the assumption  that  there  is no depletion  of the
ground state.  The last step  follows directly from statement
(ii). The pre-factor, $e^{\textit{i}I_p\textit{t}}$, represents
the phase oscillations which describes the accumulated  electron
energy in the ground state ($I_p=-E_0$ is the ionization potential
of the atomic system, with $E_0$, the ground state  energy of the
atomic system). Furthermore, the transition amplitude to the
continuum states is denoted by  $\textit{b}(\textbf{v},t)$ and it
depends both on the kinetic momentum  of the outgoing electron and
the laser pulse.  Therefore, our main task will be to derive a
general expression for the amplitude  ${b}({\bf v},t)$. In order
to do so, we substitute Eq.~(\ref{PWavef}) in Eq.~(\ref{Eq:SE})
and by considering, $\hat{H_0}|0\rangle= -I_p |0\rangle$, and
$\left[-\frac{1}{2}\nabla^2 +V(\textbf{r})\right]
|\textbf{v}\rangle = \frac{\textbf{v}^2}{2}|\textbf{v}\rangle$,
the evolution of the transition amplitude becomes:
\begin{eqnarray}
i\int{\textit{d}^3  \textbf{v} \:\dot{b}( \textbf{v},t)\: |\textbf{v} \rangle} &=& \,\, \int{\textit{d}^3 \textbf{v} \bigg(\frac{\textbf{v}^2}{2}+ I_p \bigg)\textit{b}( \textbf{v},t) |\textbf{v}}\rangle+\textbf{E}(t) \cdot  \textbf{r} |0 \rangle \nonumber \\
 & &+ \textbf{E}(t) \cdot \int{ d^3 {\bf v} \left[i{\nabla}_{\bf v}b( \textbf{v},t)+\textit{b}( \textbf{v},t){\bf r}\right]|{\bf v}\rangle}. \label{Eq:TempEq}
\end{eqnarray}

Note that we have assumed that  the electron-nucleus  interaction
is neglected once the electron  appears  at the continuum,
i.e.~$V({\bf r})|{\bf v} \rangle = 0$, which corresponds to the
statement (iii). Therefore, by multiplying Eq.~(\ref{Eq:TempEq})
by $\langle {\bf v}'|$ and after some algebra, the time variation
of the transition amplitude $\textit{b}(\textbf{v},t)$ reads:
\begin{eqnarray}
 \dot{b}( \textbf{v},t) &=&  -i\left(\frac{\textbf{v}^2}{2}+ I_p \right)\textit{b}( \textbf{v},t) +i\textbf{E}(t) \cdot {\bf d}({\bf v})\nonumber \\
& & +\,{\bf E}(t)\cdot\nabla_{\bf v} b({\bf v},t)-i \textbf{E}(t)\cdot \int{\textit{d}^3 \textbf{v}^{\prime}\: \textit{b}( \textbf{v}^{\prime},t){\bf g}( \textbf{v}, \textbf{v}^{\prime})}.
 \label{Eq:New5}
 \end{eqnarray}

The first term on the right-hand of Eq.~(\ref{Eq:New5}) represents
the  phase evolution  of the  electron in the oscillating laser
field. In the second term  we have defined the bound-free
transition dipole matrix  element as:
\begin{equation}
 - \langle \textbf{v} |\textbf{r}|0 \rangle=\textbf{d}( \textbf{v}),
   \label{Eq:dv}
\end{equation}
and  finally,  the  last  two  terms  describe the
continuum-continuum transition, $\nabla_{\bf v} b({\bf v},t)$,
without   the  influence  of  the  scattering   center,  and  by
considering  the  core  potential, $\int{\textit{d}^3
\textbf{v}^{\prime}\: \textit{b}( \textbf{v}^{\prime},t){\bf
g}({\bf v},{\bf v}')}$. Here, ${\bf g}({\bf v},{\bf v}')$, denotes
the  re-scattering  transition matrix  element, where the
potential  core plays an essential role:
\begin{eqnarray}
  \langle \textbf{v} |\textbf{r}|\textbf{v}'\rangle&= \textbf{g}(\textbf{v},\textbf{v}').
  \label{Eq:CCM1}
\end{eqnarray}

In the following, we shall describe how it is possible to compute
the transition amplitude, $b({\bf v},t)$, by applying  the  zeroth
and  first order  perturbation theory  to  the  solution  of the
partial  differential  equation  Eq.~(\ref{Eq:New5}). Therefore,
according  to this  perturbation theory,  we split  the  solution
of the  transition amplitude, $b({\bf v},t)$,
 into two parts: $b_0({\bf v},t)$ and $b_1({\bf v},t)$, i.e.~$b({\bf v},t)=b_0({\bf v},t)+b_1({\bf v},t)$.  The zeroth order of our perturbation theory $b_0({\bf v},t)$ will be called the direct term.  It describes the transition amplitude  for a laser-ionized electron that  will never re-scatter  with  the  remaining  ion-core.   On  the  other  hand,  the  first  order  term,  named re-scattered term, $b_1({\bf v},t)$, is referred to the  electron  that,  once ionized, will have a certain probability  of re-scattering  with the potential  ion-core.

\subsection*{Direct transition amplitude}

Let us consider the  process where the  electron is ionized
without probability  to return to its parent ion. This process is
modeled by the direct photoelectron  transition amplitude
$b_0({\bf v},t)$. As the  direct  ionization  process should have
a larger  probability  compared  with the re-scattering
one~\cite{Lewenstein1995},  one can neglect the last term in
Eq.~(\ref{Eq:New5}). This is what we refer to zeroth order
solution:
\begin{equation}
{\partial }_tb_0( \textbf{v},t) =-i\left(\frac{\textbf{v}^2}{2}+ I_p \right){b}_0( \textbf{v},t)  + \textit{i}\: \textbf{E}(t) \cdot  \textbf{d}( \textbf{v}) +\textbf{E}(t) \cdot \nabla_{\bf v}\textit{b}_0( \textbf{v},t).
\end{equation}

The above equation is a first-order inhomogeneous differential
equation, which is easily solved by conventional  integration
methods (see e.g.~\cite{Lev}). Therefore, the solution can be
written as:
\begin{equation}
\begin{split}
b_0( \textbf{v},t) = &i \int_0^t{\textit{d} \textit{t}^{\prime}\:\textbf{E}(t^{\prime})}\:\cdot \textbf{d}\left( \textbf{v}-\textbf{A}(t)+\textbf{A}(t^{\prime})\right)\\
&\times \exp\left(-\textit{i} \:\int_{t^{\prime}}^{t}{dt^{{\prime}{\prime}}[(\textbf{v}-\textbf{A}(t)+\textbf{A}(t^{{\prime}{\prime}}))^2/2 +I_p]}\right). \label{Eq:IntDirecTerm}
 \end{split}
\end{equation}

Here, we have considered that the electron appears in the
continuum with kinetic momentum ${\bf v}(t')={\bf v}-{\bf
A}(t)+{\bf A}(t')$ at the time $t'$, where {\bf v}~is the final
kinetic momentum  (note  that in virtue  of using atomic  units,
where the  electron  mass $m=1$, the kinetic electron momentum
${\bf p}_e$~and the electron velocity ${\bf v}$ have the same
magnitude and direction), and $\textbf{A}(t) =-\int^{t}{
\textbf{E}(t^{\prime})dt^{\prime}}$ is the vector potential of the
electromagnetic field. In particular, the vector potential at the
time when the electron appears at the continuum $t'$ is denoted by
${\bf A}(t')$ and at a certain detection time $t$, the vector
potential reads ${\bf A}(t)$. In addition, it is possible to write
Eq.~(\ref{Eq:IntDirecTerm}) as a function of the canonical
momentum ${\bf p}$, defined by ${\bf p} = {\bf v} - {\bf A}(t)$,
and therefore the probability transition amplitude for the direct
electrons simplifies to~\cite{Lewenstein1994}:
\begin{equation}
\begin{split}
b_0( \textbf{p},t) = &\textit{i} \:\int_0^t{\textit{d} \textit{t}^{\prime}\:\textbf{E}(t^{\prime})}\:\cdot \textbf{d}\left( \textbf{p}+\textbf{A}(t^{\prime})\right)\\
&\times \exp\left(-\textit{i} \:\int_{t^{\prime}}^t{d{\tilde t}\,[(\textbf{p}+\textbf{A}({\tilde t}))^2/2 +I_p]}\right).
 \end{split}
 \label{Eq:b_0}
\end{equation}

This expression is understood  as the sum of all the ionization
events which occur from the time $t'$ to $t$. Then,  the
instantaneous  transition probability  amplitude  of an  electron
at a time  $t'$, at  which it  appears  into  the  continuum  with
momentum  ${\bf v}(t')= \textbf{p}+\textbf{A}(t^{\prime})$, is
defined by the argument of the integral in Eq.~(\ref{Eq:b_0}).
Furthermore, the exponent phase factor in Eq.~(\ref{Eq:b_0})
denotes the ``semi-classical action", ${S}({\bf p},t,t^{\prime})$,
that defines a possible electron trajectory  from the birth time
$t'$ until the ``detection" one $t$~\cite{Lewenstein1995}:
 \begin{equation}
{S}({\bf p},t,t^{\prime}) = \int_{t^{\prime}}^{t}{\:d{\tilde t}\left[({\bf p}+\textbf{A}({\tilde t}))^2/2 +I_p\right]}.
\end{equation}

As our purpose is to obtain the final transition amplitude
$b_0({\bf p},t)$, the time $t$~will be fixed at the end of the
laser field, $t=t_{\rm F}$. For our calculations, we shall define
the integration time window as: $t$:$\,\,[0,t_{\rm F}]$.
Therefore, we set, ${\bf E}(0) = {\bf E}(t_{\rm F}) = {\bf 0}$, in
such a way to make sure that the  electromagnetic  field is a time
oscillating wave and  does not  have static  components. The  same
arguments  are applied to the  vector  potential  ${\bf A}(t)$. We
have defined the laser pulse envelope as
$f(t)=\sin^2(\frac{\omega_0t}{2N_c})$ where $N_c$ denotes the
number of total cycles.

\subsection*{Re-scattering transition amplitude}

In order to find a solution for the transition amplitude  of the
re-scattered photoelectrons, $b_1({\bf v},t)$,  we have considered
the re-scattering core matrix element
$\textbf{g}(\textbf{v},\textbf{v}^{\prime})$ term of
Eq.~(\ref{Eq:New5}) different than zero,
i.e.~$\textbf{g}(\textbf{v},\textbf{v}^{\prime}) \not=
\textbf{0}$. In addition,  the  first-order  perturbation theory
is applied to obtain $b_1({\bf v},t)$ by inserting the
zeroth-order solution $b_0({\bf p},t)$ in the right-hand side of
Eq.~(\ref{Eq:New5}). Then, we obtain $b_1({\bf p},t)$ as a
function of the canonical momentum ${\bf p}$ as follows:
\begin{equation}
\begin{split}
b_1( \textbf{p},t) =& -\int_0^t{\textit{d}t^{\prime}\exp{\left[-\textit{i} S({\bf p},t,t')\right]}\,{\textbf{E}(t^{\prime})\cdot}} \int_0^{t^\prime}{\textit{d} \textit{t}^{{\prime}{\prime}}}\int{\textit{d}^3\textbf{p}^{\prime} }  \textbf{g}\left(\textbf{p}+\textbf{A}(t^{\prime}),\textbf{p}^{\prime}+\textbf{A}(t^{\prime})\right)\\
 &\times \textbf{E}(t^{{\prime}{\prime}}) \cdot \textbf{d}\left( \textbf{p}^{\prime} +\textbf{A}(t^{{\prime}{\prime}})\right)\: \exp{\left[-\textit{i}  S({\bf p}',t',t'')\right]}.
  \end{split}
    \label{Eq:b_1}
\end{equation}

This last equation contains all the information about the
re-scattering  process. In particular, it is referred to the
probability  amplitude  of an emitted  electron  at  the  time
$t^{\prime\prime}$, with an amplitude given by $
\textbf{E}(t^{{\prime}{\prime}}) \cdot \textbf{d}\left(
\textbf{p}^{\prime} +\textbf{A}(t^{{\prime}{\prime}})\right)$. In
this step the electron has a kinetic momentum of ${\bf
v}^{\prime}(t'')=\textbf{p}^{\prime}+\textbf{A}(t^{{\prime}{\prime}})$.
The last factor, $\exp{\left[-\textit{i} S({\bf
p}^{\prime},t{'},t'')\right]}$, is the accumulated phase of an
electron born at the time $t^{\prime\prime}$ until it re-scatters
at time $t^\prime$. The term, ${\bf g}({\bf p}+{\bf A}(t'),{\bf
p}'+{\bf A}(t'))$, contains  the  structural matrix  element  of
the  transition continuum-continuum at  the  re- scattering  time
$t'$. At this particular moment in time,  the electron changes its
momentum  from ${\bf p}'+{\bf A}(t')$ to ${\bf p}+{\bf A}(t')$. We
stress out, however, that  the term   ${\bf g}({\bf v},{\bf v}')$
does not necessarily imply that  the electron returns to the ion
core.  In addition, $\exp\left[-\textit{i}S({\bf p},t,t')\right]$
defines the accumulated phase of the electron after the
re-scattering from the time $t'$ to the ``final" one $t$ when the
electron is ``measured" at the detector with momentum \textbf{p}.
In particular, note that  the photoelectron  spectra, $|b({\bf
p},t_{\rm F})|^2 $, is a coherent superposition of both solutions,
$b_0({\bf p},t_{\rm F})$ and $b_1({\bf p},t_{\rm F})$:
 \begin{eqnarray}
|b({\bf p},t_{\rm F})|^2 &=& |b_0({\bf p},t_{\rm F})+b_1({\bf p},t_{\rm F})|^2,\nonumber \\
&=& |b_0({\bf p},t_{\rm F})|^2 + |b_1({\bf p},t_{\rm F})|^2 + b_0({\bf p},t_{\rm F}){b_1^*}({\bf p},t_{\rm F}) + c.c. \label{Eq:ATIS}
\end{eqnarray}

So far we have formulated  a model, which describes the
photoionization process leading to two main terms,  namely, a
direct $b_0({\bf p},t_{\rm F})$ and a re-scattering $b_1({\bf
p},t_{\rm F})$ one. As the complex transition amplitude,
Eq.~(\ref{Eq:b_0}), is a ``single time integral", it can be
integrated numerically  without  major problems.  However, the
multiple  time (``2D") and momentum (``3D") integrals of the
re-scattering term, Eq.~(\ref{Eq:b_1}), present a very difficult
and demanding task from a computational perspective.  In order to
reduce the computational difficulties, and to obtain a physical
meaning of the ATI process, we shall employ the stationary phase
method  to evaluate these highly oscillatory integrals.

The fast oscillations of the momentum ${\bf p}'$ integral for the
electron re-scattering  transition amplitude, $b_1({\bf p},t)$,
suggests to use the stationary-phase approximation or the saddle
point method to solve Eq.~(\ref{Eq:b_1}). This method  is expected
to be accurate,  when both  the  $U_p$ and the $I_p$, as well as
the involved momentum ${\bf v}$ and ${\bf v}'$, are large. As the
quasi-classical action $S(\textbf{p}^{\prime},t',t'')$, is
proportional to $I_p$, $U_p$ and ${\bf v}'^2$, the phase factor,
$\exp(-\textit{iS}(\textbf{p}^{\prime},t',t'')$, oscillates very
rapidly. Then, the integral over the momentum
$\mathbf{p}^{\prime}$ of Eq.~(\ref{Eq:b_1}) tends towards  zero
except near the  extremal points   of the  phase, i.e. $
\nabla_{{\bf p}^{\prime}} \textit{S}(\textbf{p}^{\prime})={\bf
0}$. Thus, the main contributions to the momentum integral are
dominated by momenta, ${\bf p}'_s$, which satisfy the solution of
the equation: $ \nabla_{{\bf p}^{\prime}}
\textit{S}(\textbf{p}^{\prime})|_{{\bf p}'_s}={\bf 0} $. These
saddle point momenta read:
\begin{equation}
\textbf{p}'_s = -\frac{1}{\tau}\int_{t^{{\prime}{\prime}}}^{t^\prime}  { \textbf{A}(\tilde{t}) \textbf{d}\tilde{t}}.
\end{equation}

Here, $\tau=t^{\prime}-t^{\prime \prime}$ is the excursion time of
the electron in the continuum. In terms of Classical Mechanics,
these momenta roots ${\bf p}'_s$ are those corresponding to the
classical electron trajectories because the  momentum  gradient of
the  action  can be understood as the  displacement  of a
particle~\cite{Goldstein}. As the momentum gradient of the action
is null $\Delta {\bf r}=\nabla_{{\bf p }'}S({\bf p}',t',t'')={\bf
0}$, the considered electron trajectories, ${\bf r}(t)$, are for
an electron that is born at the time $t''$ at a certain position
${\bf r}(t'')={\bf r}_0$. Then, after some time $t'$ the electron
returns to the initial position ${\bf r}(t')={\bf r}_0$ with an
average momentum ${\bf p}'_s$.

Therefore, the function $ \textit{S}(\textbf{p}^{\prime},t',t'') $
can be expanded in Taylor series around the roots $\textbf{p}'_s$
and the transition amplitude for the re-scattering electrons
$b_1({\bf p},t)$ becomes:
 \begin{equation}
\begin{split}
b_1( \textbf{p},t) = & -\int_0^t{\textit{d}t^{\prime}}{e^{-\textit{i}\int_{t^\prime}^t{d{\tilde t}\left[(\textbf{p}+\textbf{A}({\tilde t}))^2/2 +I_p\right]}}\,\,\,\textbf{E}(t^{\prime})\,\, \cdot}\int_0^{t{^\prime}}{\textit{d} \textit{t}^{{\prime}{\prime}}} \textbf{g}\left(\textbf{p}+\textbf{A}(t^{\prime}),\textbf{p}'_s+\textbf{A}(t^{\prime})\right) \\
 &\times \left( \frac{\pi}{\varepsilon +{\textit{i}(t'-t'')}/{2}}  \right)^{\frac{3}{2}} \textbf{E}(t^{{\prime}{\prime}}) \cdot \textbf{d}\left( \textbf{p}_{s}^{\prime} +\textbf{A}(t^{{\prime}{\prime}})\right)\: e^{-\textit{i}\int_{t^{\prime\prime}}^{t^\prime}{d{\tilde t}\:\left[(\textbf{p}_s^{\prime}+\textbf{A}({\tilde t}))^2/2 +I_p\right]}}.
  \end{split}
  \label{Eq:B1}
  \end{equation}

Here, we have introduced a smoothing parameter, $\varepsilon$, to
avoid the divergence at the time $t'=t''$. Note that the 3D
momentum integral on ${\bf p}'$ of Eq.~(\ref{Eq:b_1}) can then be
solved by:
  \begin{equation}
\int{\textit{d}^3\textbf{p}^{\prime}  \textit{f}\:(\textbf{p}_s^{\prime}) \exp{\left(-\textit{i}\left[S(\textbf{p}_s^{\prime})+\frac{1}{2}\nabla^2_{{\bf p}'}S({\bf p}')\biggr|_{{\bf p}'_s} \cdot ({\bf p}'-{\bf p}'_s)^2\right]\right)}}   \approx \left( \frac{\pi}{\varepsilon +\frac{\textit{i}(t'-t'')}{2}}  \right)^\frac{3}{2}\textit{f}\:(\textbf{p}_s^{\prime}).
 \end{equation}

With the last equation we have substantially reduced the
dimensionality of the problem from a 5D integral to a 2D integral.
As the computing time depends on the dimensionality of the
integration problem, this reduction is extremely advantageous from
a computational viewpoint.  Moreover,  with  the  saddle  point
method  a quasi-classical  picture  for the  re-scattering
transition amplitude  is obtained  similar to the approach
described in~\cite{Corkum1994, Lewenstein1995}.

The main problem to calculate the ATI spectrum is then the
computation of the bound- free transition dipole matrix  element,
${\bf d}({\bf v})$, and  the  continuum-continuum transition
re-scattering  matrix  element ${\bf g}({\bf v},{\bf v}')$ for a
given atomic system. In the next section, we shall introduce a
short-range potential model in order  to  compute  the  transition
matrix elements and the final photoelectron  momentum distribution
analytically.

\section{Above-threshold ionization in atomic systems} 

In this section, as a test case for our  model, we chose a
non-local atomic  potential  with  the  purpose  of computing
both the  direct  and  the  re-scattering   transition
amplitudes. These  terms  involve  the  dipole  and  the
continuum-continuum matrix  elements  defined  by
Eqs.~(\ref{Eq:dv}) and (\ref{Eq:CCM1}). Then,  our  main  task
will be devoted to analytically  find the wavefunctions for the
ground  and scattering  states  of our test  potential. The
Hamiltonian, $\hat{H}(\textbf{p},\textbf{p}^{\prime}) $,  of the
atomic system in the momentum representation can be written as:
 \begin{equation}
\hat{H}(\textbf{p},\textbf{p}^{\prime}) = \frac{{\bf p}^2}{2}\delta(\textbf{p}-\textbf{p}^{\prime}) + \hat{V}(\textbf{p},\textbf{p}^{\prime}),
\end{equation}
where the first term  on the right-hand side is the kinetic energy
operator,  and the second one is the  interacting non-local
potential $\hat{V}(\bf{p},{\bf p}')$. By using such  Hamiltonian,
we write the stationary  Schr\"odinger equation as follows:
\begin{eqnarray}
\hat{H}(\textbf{p},\textbf{p}^{\prime})\Psi(\textbf{p}) &=& \int{\textit{d}^3 \textbf{p}^{\prime}\hat{H}(\textbf{p},\textbf{p}^{\prime})\Psi(\textbf{p}^{\prime})},   \nonumber\\
E\: \Psi(\textbf{p})&=&\frac{p^2}{2} \: \int{\textit{d}^3 \textbf{p}^{\prime} \delta(\textbf{p}-\textbf{p}^{\prime}) \Psi(\textbf{p}^{\prime})}  -\gamma \phi(\textbf{p})\: \int{\textit{d}^3 \textbf{p}^{\prime}\phi(\textbf{p}^{\prime})\Psi(\textbf{p}^{\prime})},
\end{eqnarray}
where $E$ denotes the energy of the wavefunction $\Psi({\bf p})$.
Note that we have defined the non-local potential as
$\hat{V}(\textbf{p},\textbf{p}^{\prime})= -\gamma
\phi(\textbf{p})\: \phi(\textbf{p}^{\prime})$, which describes the
attraction between the electron  and  the
nucleus~\cite{Lewenstein1995}. This  potential  has been chosen
such that it assures analytical  solutions of the continuum  or
scattering  states,  i.e.~for states  with energies $E>0$. Note
that the  ground  state  can  also be  calculated  analytically.
$\gamma$ is understood as a screening parameter and $\phi({\bf
p})$ is an auxiliary function defined by:
 \begin{equation}
 \phi(\textbf{p}) = \frac{1}{\sqrt{{\bf p}^2 +\Gamma^2}},
 \label{Eq.Potential}
\end{equation}
where the parameter $\Gamma $ is a constant related with the shape
of the ground state.  In order to analytically  obtain  the
ground  state, $\Psi_0({\bf p})$, we solve the stationary
Schr\"odinger equation in the momentum representation:
\begin{equation}
\frac{p^2}{2}\: \Psi_0(\textbf{p})  -\frac{\gamma}{\sqrt{p^2 + \Gamma ^2}} \int{\frac{\textit{d}^3 \textbf{p}^{\prime} \Psi_0(\textbf{p}^{\prime})}   {\sqrt{{p^{\prime}}^2 + \Gamma ^2}}  } = E_0\: \Psi_0(\textbf{p}),
\label{Eq:Sch2}
\end{equation}
where the parameter $\gamma$ is related to the ionization
potential, $I_p$, of the atomic species under study. To solve
Eq.~(\ref{Eq:Sch2}) we consider $\check{\varphi} =
\int{\frac{\textit{d}^3 \textbf{p}^{\prime}
\Psi(\textbf{p}^{\prime})} {\sqrt{{p^{\prime}}^2 + \Gamma ^2}} }$
as a new parameter and write the eigenenergy~$E_0=-I_p$.
Therefore, the final solution reads:
  \begin{eqnarray}
 \Psi_0(\textbf{p}) &=& \frac{\mathcal{N}}{\sqrt{(p^2 + \Gamma ^2)}(\frac{p^2}{2} +I_p)}
 \label{Eq:WF1}
\end{eqnarray}
where, $\mathcal{N}=\gamma\: \check{\varphi}$ denotes a
normalization constant. Dividing the last formula by $ \sqrt{p^2 +
\Gamma ^2}$ and taking the volume integral on $\textbf{p}$, we
obtain:
\begin{equation}
\check{\varphi} = \gamma \check{\varphi}  \int{\frac{\textit{d}^3 \textbf{p}}{({p}^2 + \Gamma ^2) (\frac{p^2}{2} + I_p)} }. \label{Eq:Ncond}
\end{equation}

The solution of the last integral in Eq.~(\ref{Eq:Ncond}) gives us
the relation between the parameters $I_p$, $\Gamma$ and $\gamma$:
\begin{eqnarray}
 \gamma  \int_0^{\pi}{\textit{d}\theta} \int_0^{2\pi}{\textit{d}\varphi \: \sin\varphi} \int_0^{\infty}{\frac{\textit{d} {p} \:{ p}^2}{({p}^2 + \Gamma ^2) (\frac{p^2}{2} +I_p)} }&=&1, \nonumber \\
\frac{\gamma\: 4\pi^2}{\Gamma + \sqrt{2I_p}} &=&1.
\label{Eq:gamma}
\end{eqnarray}

This formula allows us to control the parameters $\Gamma$ or
$\gamma$, in such a way as to match the $I_p$ of the atomic
system. Furthermore, by using the normalization condition for the
bound states, we calculate  the normalization  constant,
$\mathcal{N}$, as well as the analytical ground wave function
$\Psi_{0}({\bf p})$. This normalization factor reads:
 \begin{equation}
 \mathcal{N}^2 = \frac{\sqrt{2I_p}\bigg( \Gamma +  \sqrt{2I_p} \bigg)^2}{4\pi^2}.
 \end{equation}

So far we have obtained, analytically,  the  ground  state  of our
non-local potential model. This ground state  will allow us to
calculate the bound-free transition dipole matrix  element by
using Eq.~(\ref{Eq:dv}). The free or continuum  state  is
approximated as a plane wave of a given momentum, ${\bf p}_0$, and
therefore  the  bound-free  transition dipole matrix  in the
momentum representation reads:
\begin{equation}
\textbf{d}( \textbf{p}_0) =\textit{i}\int{ \Psi_0(\textbf{p}')\nabla_{\textbf{p}'}\delta(\textbf{p}'-\textbf{p}_0)\:  \textit{d}^3 \textbf{p}' }.
\end{equation}

By employing properties of the Dirac delta distribution,
$\textbf{d}( \textbf{p}_0)$ is computed via $\textbf{d}(
\textbf{p}_0)=  -\textit{i}\nabla_{\textbf{p}'}
\Psi_0(\textbf{p}')\big\rvert_{{\bf p}_0}$. After some elementary
algebra, we obtain  the transition dipole matrix :
\begin{eqnarray}
\textbf{d}( \textbf{p}_0) &= -\textit{i}{\nabla_{{\bf p}'}} \left(\frac{\mathcal{N}}{(p'^2 + \Gamma^2)^{\frac{1}{2}}(\frac{p'^2}{2} + I_p)} \right)\biggr\rvert_{{\bf p}_0},\nonumber\\
  &= \textit{i}\mathcal{N} \textbf{p}_0\,\, \frac{(p_0^2 + \Gamma^2) + (\frac{p_0^2}{2} + I_p)}{(p_0^2 + \Gamma^2)^{\frac{3}{2}}(\frac{p_0^2}{2} + I_p)^2}.
 \label{Eq.dp1}
 \end{eqnarray}

The  second important quantity to be calculated  before evaluating
the whole transition amplitude  $b({\bf p},t)$ is the transition
continuum-continuum matrix  element, ${\bf g}({\bf p},{\bf p'})$.
Hence, we need to find the scattering  or continuum wave functions
of our model potential.  Next, we shall calculate the scattering
states by analytically  solving the time independent Schr\"odinger
equation  in the momentum  representation for positive energies.

\subsection*{Scattering waves and continuum-continuum transition matrix element}

Let us consider the scattering wave, $\Psi_{{\bf p}_0}({\bf p})$,
with asymptotic momentum ${\bf p}_0$, as a coherent superposition
of a plane wave and an extra correction $\delta\Psi_{{\bf
p}_0}({\bf p})$:
\begin{eqnarray}
\Psi_{\textbf{p}_0}(\textbf{p})  = \delta(\textbf{p}-\textbf{p}_0) +  \delta\Psi_{{\bf p}_0}(\textbf{p}).
 \end{eqnarray}

 This state has an energy $E={{\bf p}_0^2}/2$. Then, the Schr\"odinger equation in momentum representation reads:
\begin{eqnarray}
\frac{p_0^2}{2}\:\Psi_{\textbf{p}0}(\textbf{p}) &=&\frac{{\ p}^2}{2}\: \Psi_{\textbf{p}0}(\textbf{p})  -\frac{\gamma}{\sqrt{p^2 + \Gamma ^2}} \int{\frac{\textit{d}^3 \textbf{p}^{\prime} \: \Psi_{\textbf{p}0}(\textbf{p})}   {\sqrt{{p^{\prime}}^2 + \Gamma ^2}}  }, \nonumber\\
 \bigg(\frac{{ p}^2}{2} - \frac{p_0^2}{2} \bigg)\ \delta\Psi_{{\bf p}_0}(\textbf{p})  &=& \frac{\gamma}{\sqrt{p^2 + \Gamma ^2} \sqrt{{p}_0^2 + \Gamma ^2}}+ \frac{\gamma}{\sqrt{p^2 + \Gamma ^2}} \int{\frac{\textit{d}^3 \textbf{p}^{\prime} \: \delta\Psi_{{\bf p}_0}(\textbf{p})}   {\sqrt{{p^{\prime}}^2 + \Gamma ^2}}  }.
\end{eqnarray}

To analytically solve the last equation, we apply elementary
algebra and the following Dirac delta distribution properties:
$(\frac{p^2}{2} -
\frac{p_0^2}{2})\:\delta(\textbf{p}-\textbf{p}_0) = 0$, and
$\int{\frac{\textit{d}^3 \textbf{p}^{\prime} \:
\delta(\textbf{p}-\textbf{p}_0)} {\sqrt{{p^{\prime}}^2 + \Gamma
^2}} } = \frac{1}{\sqrt{{p}_0^2 + \Gamma ^2}}$.  Finally, the
correction $\delta \Psi_{{\bf p}_0}$, results:
\begin{eqnarray}
\delta\Psi_{{\bf p}_0}(\textbf{p})  = \frac{B({\bf p}_0)}{\sqrt{p^2+ \Gamma ^2}\bigg(p_0^2 - p^2 +\textit{i}\epsilon \bigg) }.
\label{Eq:SCorrection0}
 \end{eqnarray}
Here, $\epsilon$, is a smooth parameter to avoid the divergence at
${\bf p}={\bf p}_0$ and $B({\bf p}_0)$ is a constant, which
depends on the asymptotic momentum ${\bf p}_0$. The constant
$B({\bf p}_0)$ is defined by:
\begin{eqnarray}
B({\bf p}_0)= -2\gamma \bigg[\frac{1}{\sqrt{p_0^2+ \Gamma ^2}} + \check{\varphi}' \bigg].
\end{eqnarray}
 where  $\check{\varphi}' = \int{\frac{\textit{d}^3 \textbf{p}^{\prime} \: \delta\Psi(\textbf{p}^{\prime})}   {\sqrt{{p^{\prime}}^2 + \Gamma ^2}} }$.
In order to obtain $B({\bf p}_0)$, we proceed analogously to our procedure for obtaining  Eq.~(\ref{Eq:Ncond}). Consequently, for Eq.~(\ref{Eq:SCorrection0}) we obtain the same quantity, $\check{\varphi}'$, on both the left and right-hand sides. These factors cancel each other and the constant $B(\textbf{p}_0)$ reads:
\begin{eqnarray}
B(\textbf{p}_0)  = \frac{2\gamma}{(p_0^2 +\Gamma^2)^{\frac{1}{2}}}\bigg(1- \frac{4\pi^2\textit{i}\gamma}{| \textit{p}_0| +\textit{i}\Gamma }\bigg)^{-1}.
 \end{eqnarray}

Finally, the scattering wave functions can be written as:
\begin{eqnarray}
\Psi_{\textbf{p}_0}(\textbf{p})  = \delta(\textbf{p}-\textbf{p}_0) +  \frac{B({\bf p}_0)}{\sqrt{p^2 + \Gamma ^2}\bigg(p_0^2 - p^2 +\textit{i}\epsilon \bigg)}.
\label{Eq:RescM}
 \end{eqnarray}

 The later equation tells us that the correction, $\delta\Psi_{{\bf p}_0}({\bf p}) $, to the plane wave is a function of the parameters of the atomic potential, $\Gamma$ and $\gamma$.  Therefore, the re-scattering process will depend on the shape of the potential. However, in the limit when the momentum ${\bf p}_0$ goes to infinity this correction term vanishes, i.e.~$\lim_{{\bf p}_0\to\infty}\delta\Psi_{{\bf p}_0}({\bf p}) =0$, and then the atomic potential does not play
any role in the re-scattering process.

 \subsection*{Continuum-continuum transition matrix element} 

Let us consider the scattering waves obtained in
Eq.~(\ref{Eq:RescM}) and evaluate the continuum-continuum
transition matrix element of Eq.~(\ref{Eq:CCM1}), i.e.
 \begin{eqnarray}
\textbf{g}(\textbf{p}_1, \textbf{p}_2) &=&\textit{i}\int _{-\infty}^{+\infty}{\Psi_{\textbf{p}1}^*(\textbf{p})  \, \nabla_{\textbf{p}} \, \Psi_{\textbf{p}2}(\textbf{p})\textit{d}^3 \textbf{p}},\nonumber\\
 &=&\textit{i}\int _{-\infty}^{+\infty}{\Bigg[\delta(\textbf{p}-\textbf{p}_1) +  \frac{B^*(\textbf{p}_1)}{(p^2 + \Gamma ^2)^{\frac{1}{2}}(p_1^2 - p^2 -\textit{i}\epsilon)}\Bigg]} \,\nonumber\\
 & &\times \nabla_{\textbf{p}}\Bigg[\delta(\textbf{p}-\textbf{p}_2) +  \frac{B(\textbf{p}_2)}{(p^2 + \Gamma ^2)^{\frac{1}{2}}(p_2^2 - p^2 +\textit{i}\epsilon)}\Bigg]\textit{d}^3 \textbf{p}.
\end{eqnarray}

As the first-order perturbation theory has been considered along
our derivations, all quadratic or superior terms in $\gamma$, e.g.
$B^*(\textbf{p}_1)B(\textbf{p}_2)$, are neglected. Therefore, we
obtain:
\begin{eqnarray}
\textbf{g}(\textbf{p}_1, \textbf{p}_2) &=& \textit{i}\,B(\textbf{p}_2)\,\int_{-\infty}^{+\infty}{  \delta(\textbf{p}-\textbf{p}_1)  \nabla_{\textbf{p}} \frac{1}{(p^2 + \Gamma ^2)^{\frac{1}{2}}(p_2^2 - p^2 -\textit{i}\epsilon)}   } \textit{d}^3 \textbf{p}  \nonumber \\
 &&+   \textit{i}\,B^*(\textbf{p}_1)\,\int_{-\infty}^{+\infty}{ \frac{1} {(p^2 + \Gamma ^2)^{\frac{1}{2}}(p_1^2 - p^2 +\textit{i}\epsilon)}   \nabla_{\textbf{p}} \delta(\textbf{p}-\textbf{p}_2)  }  \textit{d}^3 \textbf{p}. \label{Eq:TransitionElement}
 \end{eqnarray}
 The last momentum integrals are solved by applying the same Dirac
delta distribution property used in Eq.~(\ref{Eq.dp1}) and, after
some algebra,  the  transition matrix  continuum-continuum element
for our model potential reads:
\begin{eqnarray}
 \textbf{g}(\textbf{p}_1, \textbf{p}_2) &=\textit{i}B(\textbf{p}_2)\textbf{p}_1\left[\frac{3p_1^2 - p_2^2 +2\Gamma+\textit{i}\epsilon}{(p_1^2 +\Gamma ^2)^{\frac{3}{2}}(p_2^2 - p_1^2 -\textit{i}\epsilon)^2}\right] -\textit{i}B^*(\textbf{p}_1)\textbf{p}_2\left[\frac{3p_2^2 - p_1^2 +2\Gamma-\textit{i}\epsilon}{(p_2^2 +\Gamma ^2)^{\frac{3}{2}}(p_1^2 - p_2^2 +\textit{i}\epsilon)^2}\right].\label{Eq:RSE}
\end{eqnarray}

At this point, we have obtained  all the required  elements  to
evaluate  both  the  direct  and the re-scattering  transition
amplitude  terms defined according to Eqs.~(\ref{Eq:b_0})
and~(\ref{Eq:B1}). The developed model is an alternative way to
describe the ATI process mediated  by a strong laser pulse.  The
method is physically intuitive,  and can be understood  on the
basis of a quasi-classical picture, i.e.~electron trajectories.
This is the main difference of our approach in comparison to the
numerical solution of the TDSE, whose physical interpretation is,
in spite of its accuracy, frequently challenging.  The main
advantage  of the proposed model is that  Eqs.~(\ref{Eq:b_0}) and
(\ref{Eq:B1}) give a clear physical understanding of the ATI
process and provide rich information  about  both  the laser field
and the atomic target which are encoded into  the  complex
transition amplitude   $b({\bf p},t)=b_0({\bf p},t)+b_1({\bf
p},t)$. The  exact analytical  solutions  of those  direct  and
re-scattering   transition amplitudes  are however not nontrivial
to obtain if no approximations are considered.  In particular, for
the re-scattering photoelectrons,  the solution is even more
complex and depends, generally, of the laser electric field shape.

In the  next  Section,  we numerically  integrate  both terms,
i.e. $b_0({\bf p},t)$ and $b_1({\bf p},t)$, for the non-local
potential  and compare those results to the numerical solution of
the TDSE.

\section{Results and discussion}

The numerical integration of Eqs.~(\ref{Eq:b_0}) and~(\ref{Eq:B1})
has been performed by employing a rectangular rule with dedicated
emphasis on the convergence of the results.  As the final momentum
distribution Eq.~(\ref{Eq:ATIS}) is ``locally" independent of the
momentum ${\bf p}$, i.e. $|b({\bf p},t)|^2$ can be computed
concurrently for a given set of ${\bf p}$ values, we have
optimized the calculation of the whole transition amplitude,
$|b({\bf p},t)|^2$, by using the OpenMP parallel
package~\cite{OpenMPBook}. The final momentum photoelectron
distribution, $|b({\bf p},t)|^2$, is computed both in a
1D-momentum line along~$p_z$, and in a 2D-momentum plane
$(p_y,\,p_z)$. We shall compare these  results with the numerical
solutions of the TDSE  in one (1D) and two (2D) spatial
dimensions, respectively.

In case of the 1D calculations for the ATI spectra, the momentum
grid was symmetrically defined with a length  of  ${\rm
L}_{p_z}=4.0$~a.u., and a step size of $\delta p_z=0.02$~a.u. The
parameters of the  non-local potential  are fixed to  $\Gamma=1$
and $\gamma=38$ a.u., in such a way as to match the ionization
potential of the hydrogen atom, $I_p=0.5$~a.u. Note that several
values of $\Gamma$ and $\gamma$ can be employed to obtain the same
$I_p$. Therefore, these parameters are chosen to match the ground
state  wave function, Eq.~(\ref{Eq:WF1}), of our SR potential
model with the shape of the ground state wave function of an
actual hydrogen atom. We use in our simulations an ultrashort
laser pulse with central frequency $\omega_0=0.057$~a.u.
(wavelength $\lambda=800$ nm, photon energy, $1.55$~eV), peak
intensity $I_0= 1\times10^{14}$~W$\,\cdot$\,cm$^{-2}$, with a
$\sin^2$ envelope shape with $N_c=4$ total cycles (this
corresponds to a full-width at half-maximum FWHM~$=2.67$ fs) and a
CEP $\phi_{0}=0$~rad. The time step is fixed to $\delta t=
0.2$~a.u., and the numerical integration time window is
$t$:~$[0,t_{\rm F}]$, where $t_{\rm F}= N_cT_0\approx 11$ fs and
$T_0=2\pi/\omega_0$ denote the final ``detection" time and the
cycle period of the laser field, respectively.

\begin{figure}[htb]
            \subfigure[~Full ATI photoelectron  spectrum and interference term (inset)]{\includegraphics [width=0.46\textwidth] {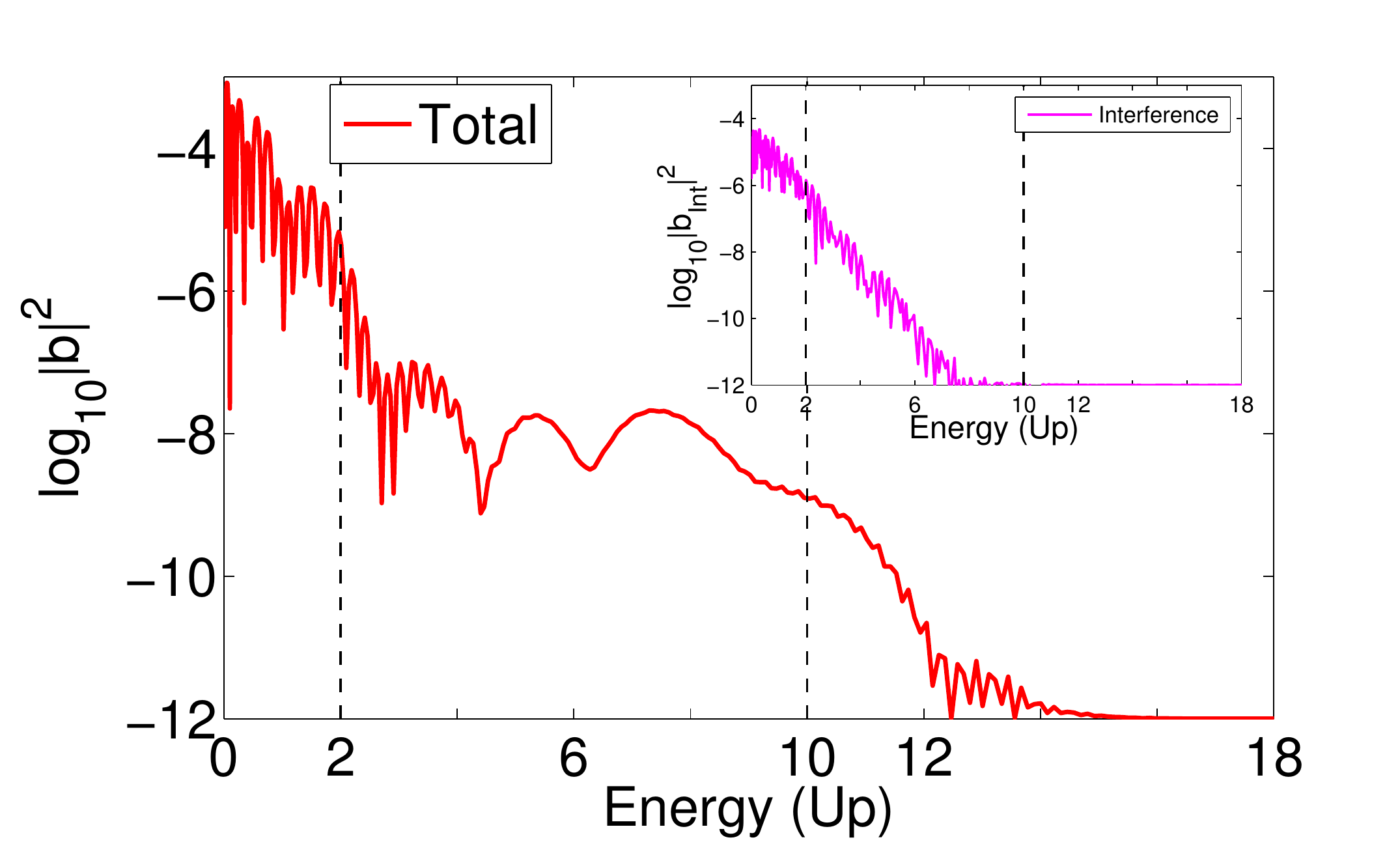}}
                         \subfigure[~Direct and re-scattering photoelectron spectra]{\includegraphics[width=0.445\textwidth]{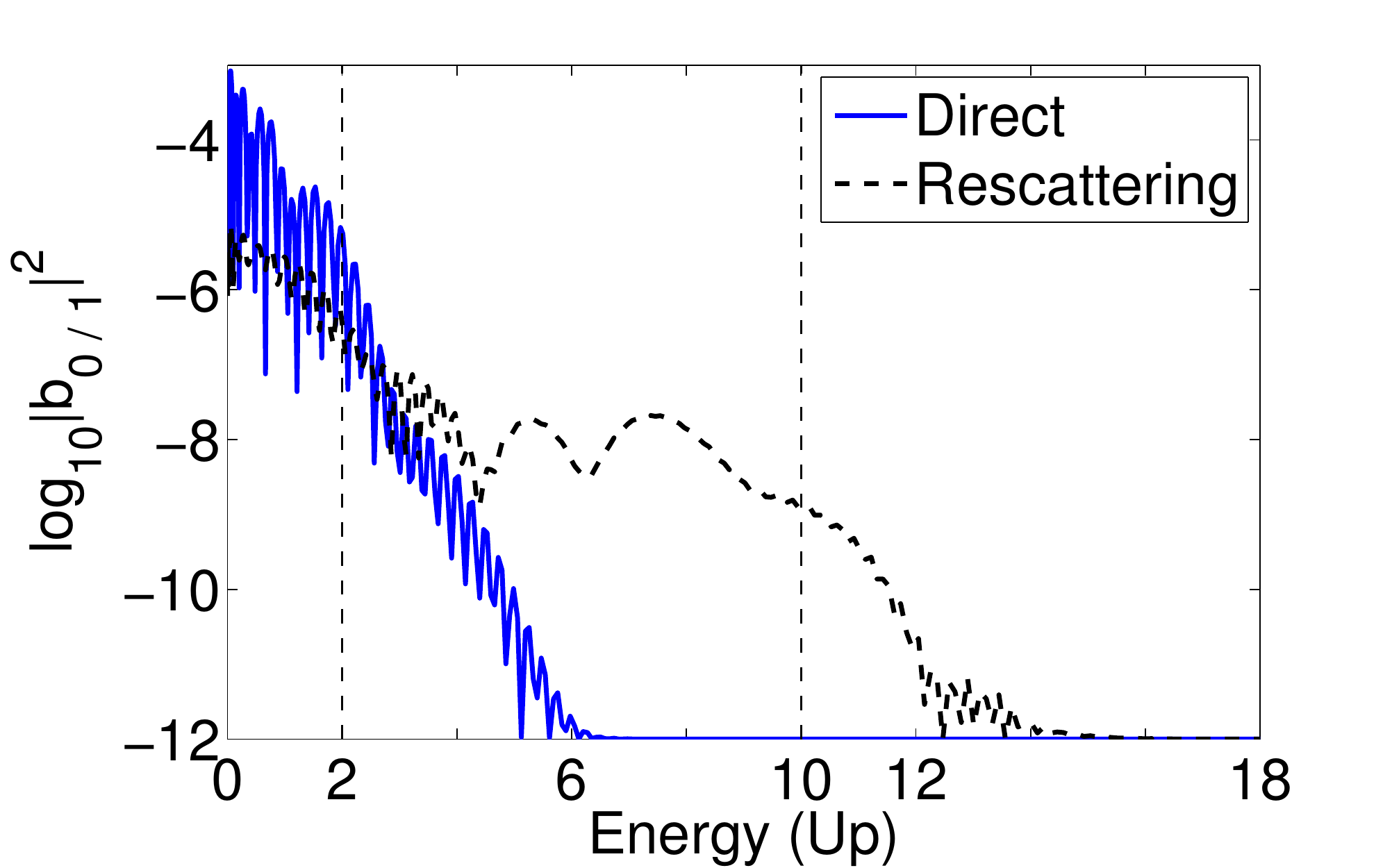}}

                    \caption{(color online) Photoelectron ATI spectra (in logarithmic scale) as a function of the ponderomotive electron energy $U_p$ computed by using our quasi-classical model and for each one of the transition terms: (a) Total photoelectron spectra, Eq.~(\ref{Eq:ATIS}), (red line) with the interference term in the inset (magenta line). (b) Direct photoelectron spectrum $|b_0({p}_z,t_{\rm F})|^2$ and re-scattering photoelectron spectrum $|b_1({p}_z,t_{\rm F})|^2$ are depicted in blue solid and black dashed lines, respectively. The vertical dashed lines correspond to the classical $2U_p$ and $10 U_p$ cutoffs (see the text for details).
                 }                  \label{Fig:TermsDRS}
    \end{figure}

Figure~\ref{Fig:TermsDRS} shows the final photoelectron
distribution or ATI spectra, in logarithmic scale, as a function
of the ponderomotive energy, $U_p$, for electrons with positive
momenta along the $p_z$-direction. Fig.~\ref{Fig:TermsDRS}(a)
depicts the total contribution, Eq.~(\ref{Eq:ATIS}), meanwhile
Fig.~\ref{Fig:TermsDRS}(b) shows the contribution of both the
direct $|b_0({\bf p},t)|^2$ and re-scattering terms $|b_1({\bf
p},t)|^2$. For completeness, the interference term, $b_{\rm
Int}({\bf p},t) = b^*_{0}({\bf p},t)\:b_{1}({\bf p},t) +
b_{0}({\bf p},t)\:b^*_{1}({\bf p},t)$ is included as an inset of
Fig.~\ref{Fig:TermsDRS}(a). The first clear observation is that
each term contributes to different regions of the photoelectron
spectra, i.e.~for electron energies $E_{p_z}\lesssim3 U_p$ the
direct term $|b_0({\bf p},t)|^2$ dominates the spectrum and, on
the contrary, it is the re-scattering term, $|b_1({\bf p},t)|^2$
the one that prevails in the high-energy electron region. In
addition, we observe that the interference term follows the trend
of the direct one (see the inset of Fig.~\ref{Fig:TermsDRS}(a))
and does not play any role for electron energies $E_{p_z}\gtrsim5
U_p$. We shall see next that both direct and re-scattering terms
are needed in order to adequately describe the ATI process.

\begin{figure}[htb]
            \subfigure[$\,$SFA Calculations]{ \includegraphics [width=0.45\textwidth] {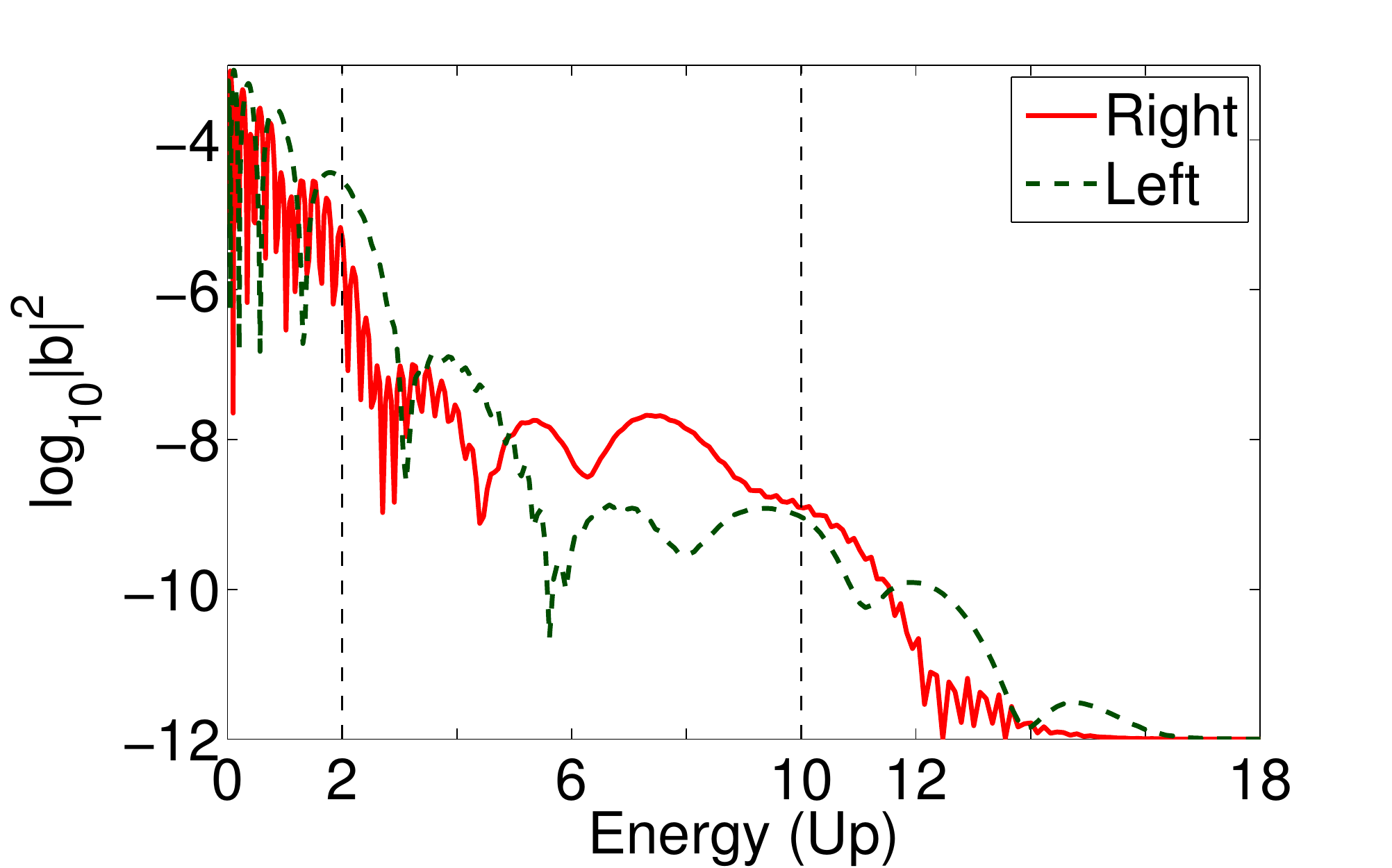}}
                         \subfigure[$\,$TDSE Calculations]{\includegraphics[width=0.45\textwidth]{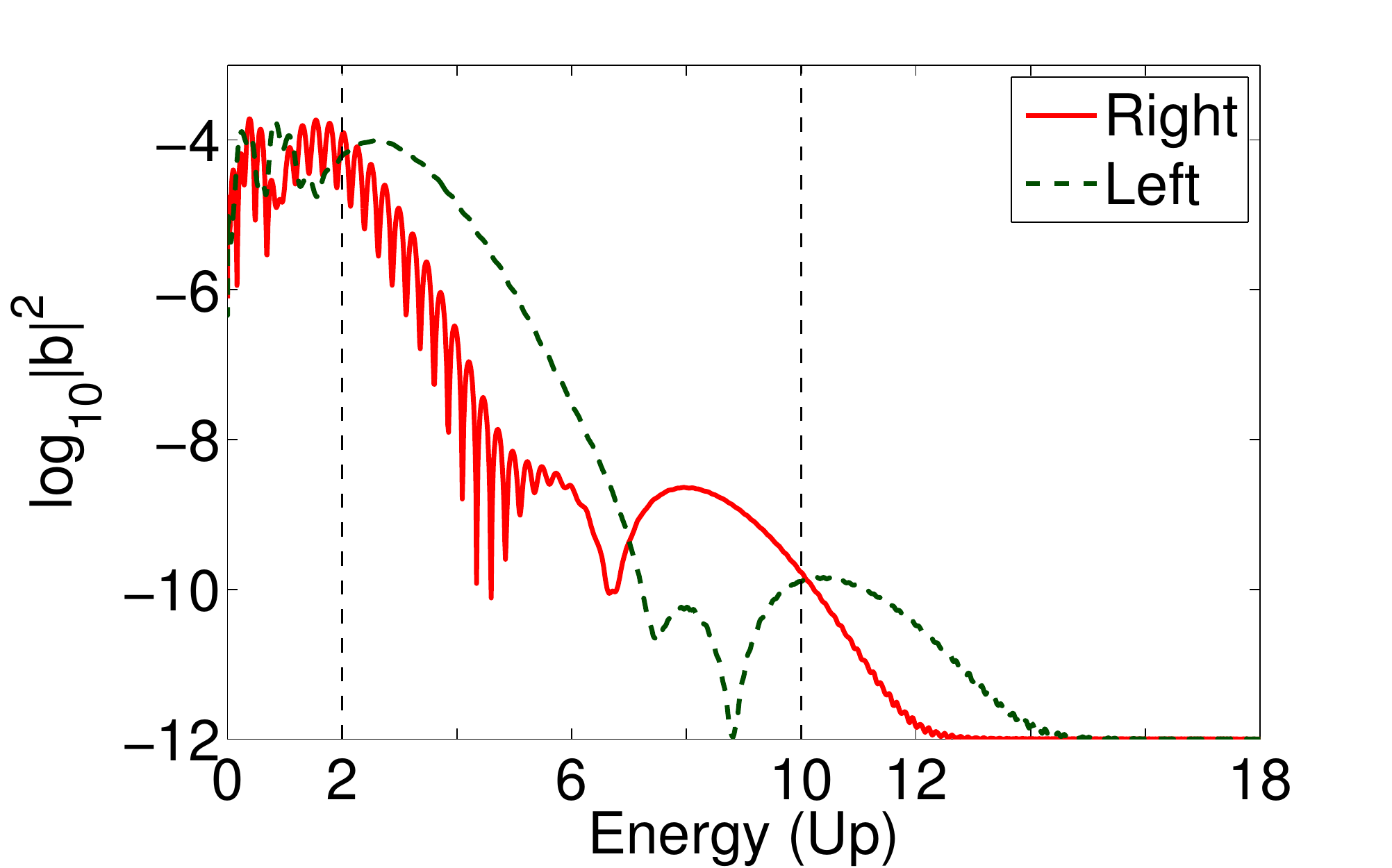}}
                    \caption{(color online) Comparison of the ATI spectra for an hydrogen atom. (a) Photoelectron energy distribution (in logarithmic scale) for the emitted electrons with negative (green dark line)
                 and positive (red line) momentum
                 obtained by the integration of our derived
                 full transition amplitude $|b({p_z},t_{\rm F})|^2$. (b) The same
                 as in (a) but computed by the numerical solution of the TDSE in 1D. The vertical dashed lines correspond to the classical $2U_p$ and $10 U_p$ cutoffs (see the text for details).
                 }                  \label{Fig:1D}
    \end{figure}

To confirm that our model is able to capture the left-right
asymmetry~\cite{Milosevic2006}, in Fig.~\ref{Fig:1D} we compute
ATI  spectra  for electrons  with  positive  and  negative
momenta  along the $p_z$-direction. Fig.~\ref{Fig:1D}(a) shows the
results  of our quasi-classical model, meanwhile in
Fig.~\ref{Fig:1D}(b) the TDSE in 1D is used. The photoelectrons
with negative ({\em positive}) momentum are conventionally named
left ({\em right}) electrons and correspondingly the photoelectron
spectra associated are labeled by $|b_{\rm L}(p_z,\phi_0)|^2$ and
$|b_{\rm R}(p_z,\phi_0)|^2$, respectively.

The photoelectron spectra computed by using the numerical solution
of the TDSE in 1D, Fig.~\ref{Fig:1D}(b),  allow us to evaluate the
accuracy  of our quasi-classical ATI model.  The numerical
integration of the TDSE is performed by using the Split-Spectral
Operator  algorithm~\cite{SplitOperator} and we use the
FFTW~\cite{fftw} to evaluate the kinetic energy operator of our
Hamiltonian $\hat{H}=\frac{\hat{p}_z^2}{2} + \hat{V}(z) + zE(t)$.
For the present numerical solution of the TDSE, we have fixed the
position grid step to $\delta z = 0.2$ a.u., with a total number
of points ${\rm N}_z = 17 000$. The ground state  is computed  via
imaginary  time propagation  with a time step of~$\delta
t=-0.02\,i$ and the soft-core Coulomb potential is given by: $V(z)
= - \frac{1}{\sqrt{ z^2 + a}}$. The parameter~$a=2$~a.u., is
chosen in such that the ground state yields the ionization
potential of the hydrogen atom, i.e.~$I_p= 0.5$~a.u.

The strong-field laser-matter interaction is simulated by evolving
the ground state wave function in real time, with a time step of
$\delta t=0.02$~a.u., and under the action of both the atomic
potential  and the laser field. The laser pulse parameters are the
same as those used to compute the results of
Fig.~\ref{Fig:TermsDRS}. At the end of the laser field $t_{\rm
F}$, when the electric field is zero, we compute the final
photoelectron energy-momentum distribution $|b_{\rm
TDSE}(p_z,t_{\rm F})|^2$, by projecting the ``free" electron wave
packet, $\Psi_c(z,t_{\rm F})$, over plane waves. The wave packet
$\Psi_c(z,t_{\rm F})$, is computed by smoothly masking the bound
states from the entire wave function function $\Psi(z,t_{\rm F})$
via: $\Psi_c(z,t_{\rm F})=h(z)\Psi(z,t_{\rm F})$, where,
$h(z)=\exp(-(\frac{z-z_0}{\sigma})^2)$ is a gaussian filter.

Figure~\ref{Fig:1D} demonstrates good qualitative agreement
between the photoelectron spectra calculated with our
quasi-classical model and those obtained  by the  numerical
solution  of the  TDSE in 1D. The left-right photoelectron spectra
show the expected two cutoffs defined by $2U_p$ and $10U_p$ (black
dashed lines) which are present in the ATI
process~\cite{Lewenstein1995,Milosevic2006}. This shows that our
approach is a reliable alternative for the  calculation  of ATI
spectra. Our model furthermore captures the left-right dependence
of the emitted photoelectrons as shown in Fig.~\ref{Fig:1D}(a),
and in comparison with the TDSE shown in Fig.~\ref{Fig:1D}(b). The
ability to capture this dependence and its features is especially
important for applications to methods such as LIED which relies on
large momentum transfers and backscattered electron distributions.
For instance, photoelectrons ejected towards the left differ
substantially from those emitted  to the right for the case when a
few-cycle driving pulse is used. According to the quasi-classical
analysis of  Section~\ref{cap:Background}, one can then infers
that electron trajectories emitted  towards the right have larger
probability to perform backward re-scattering  with the ionic core
than the electrons emitted towards the
left~\cite{NatPaulus2001,Milosevic2006}. This behavior is clearly
reproduced by both models shown in Fig.~\ref{Fig:1D} and it is the
basis for the stereo ATI technique developed by Paulus et
al.~\cite{NatPaulus2001}.

Since our model is capable of capturing the general CEP dependence
we turn to a more detailed investigation on whether our model can
reproduce detailed CEP dependence by computing the ATI spectra as
a function of the absolute laser phase $\phi_0$. Henceforth, we
define the left-right asymmetry $\mathcal{A}({ p_z},\phi_0)$ as
visibility:
\begin{eqnarray}
\mathcal{A}({ p_z},\phi_0)= \frac{|b_{\rm L}({ p_z},\phi_0)|^2 - |b_{\rm R}({ p}_z,\phi_0)|^2}{|b_{\rm R}({ p}_z,\phi_0)|^2 + |b_{\rm L}({ p}_z,\phi_0)|^2}. \label{Eq:Asymmetry}
 \end{eqnarray}

\begin{figure}[htb]
 \centering
            \includegraphics[width=0.78\textwidth] {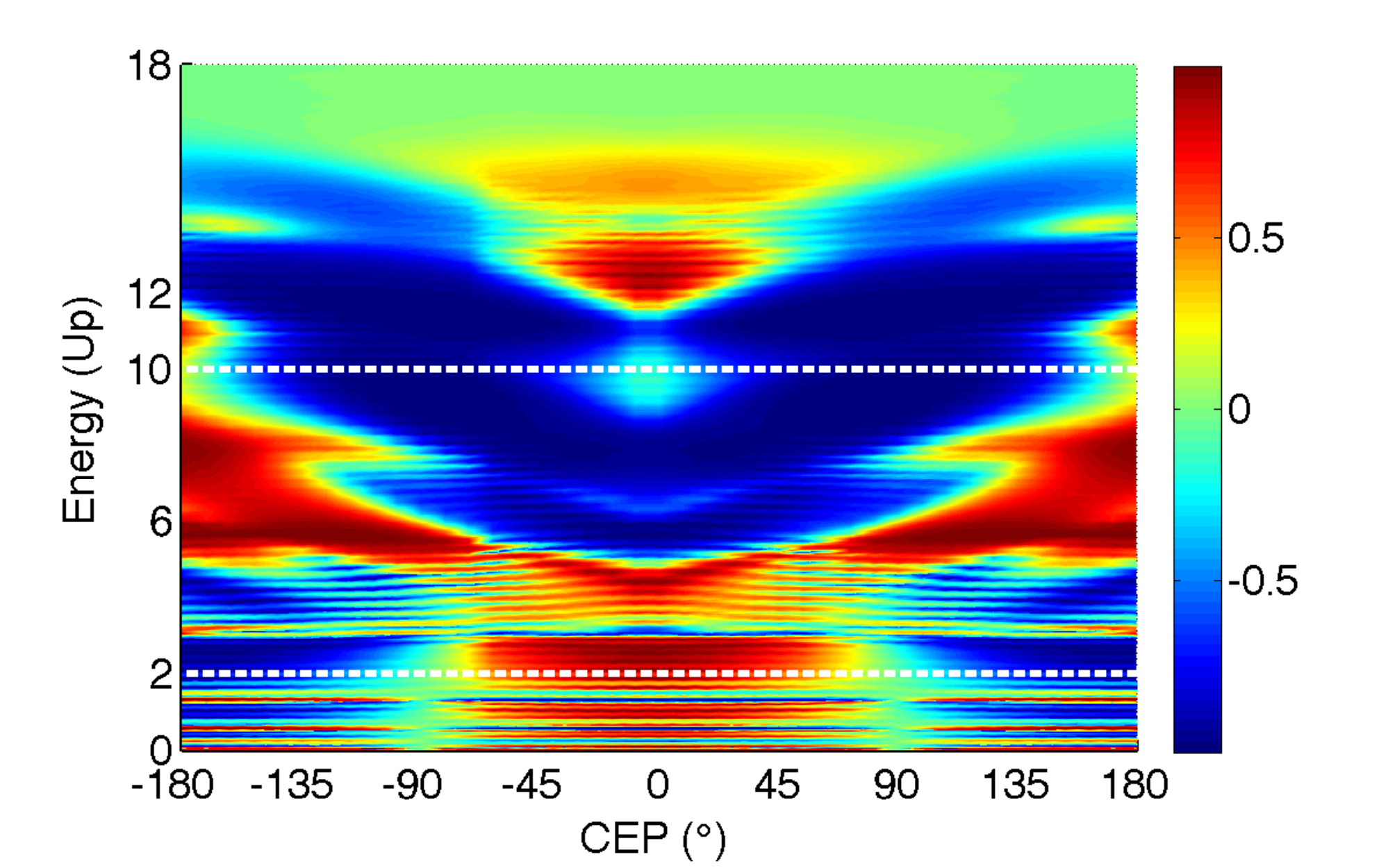}
         \caption{(color online) Asymmetry of the photoelectron energy distribution $\mathcal{A}(p_z,\phi_0)$ as a function of the CEP. The horizontal dashed white lines denote the $2U_p$ and $10U_p$ cutoffs rule for the direct and
         re-scattering photoelectrons, respectively. The laser pulse and the atomic parameters are the same that those used in Fig.~\ref{Fig:1D}.}
    \label{Fig:1DAsimetA}
\end{figure}

 We compute the ATI spectra for a set of CEP values between $\phi_0=\mp180\degree$, and
 evaluate the asymmetry $\mathcal{A}({ p_z},\phi_0)$ of Eq.~(\ref{Eq:Asymmetry}). The results are shown in Fig.~\ref{Fig:1DAsimetA}.
Our calculated asymmetry $\mathcal{A}({ p_z},\phi_0)$ shows a
clear dependence on the absolute phase $\phi_0$, of the laser
pulse. For instance, when the CEP is $\phi_0=\mp$ 90\degree, the
photoelectron spectra  show a left-right symmetry,  which is
clearly visible in the energy region between $0$ and $4U_p$ (see
Fig.~\ref{Fig:1DAsimetA}). This symmetry can be attributed to the
direct term $\textit{b}_0$, which dominates  the  photoelectron
spectra  at  lower energies and is  a consequence of the symmetry
of the  electric field with respect  to the  envelope maximum. On
the other hand,  and as we shall see later, the high-energy
re-scattered  electrons do not follow this symmetry.

\begin{figure}[htb]
            \subfigure[]{\includegraphics[width=0.48\textwidth] {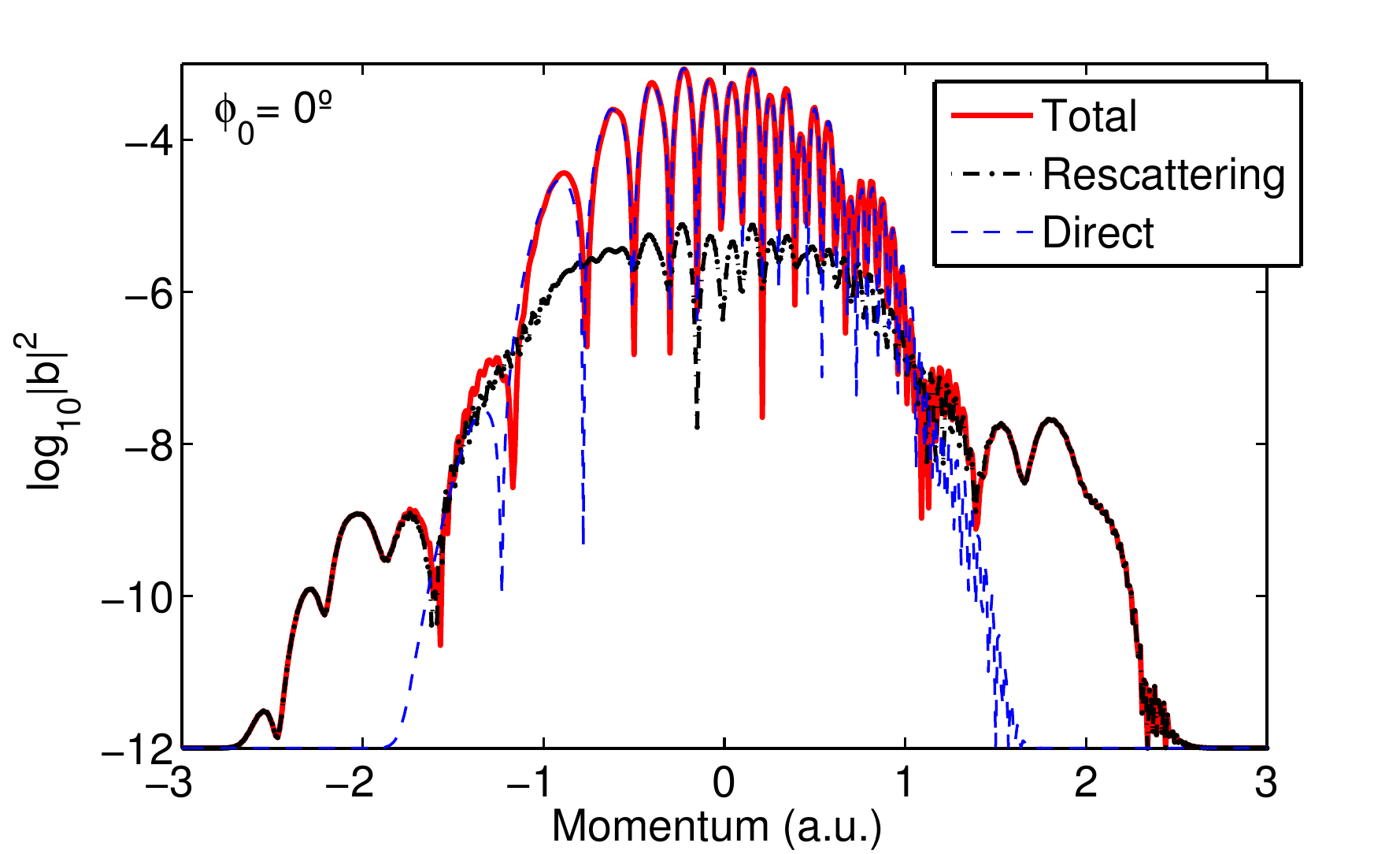}}
     \subfigure[]{\includegraphics[width=0.48\textwidth]{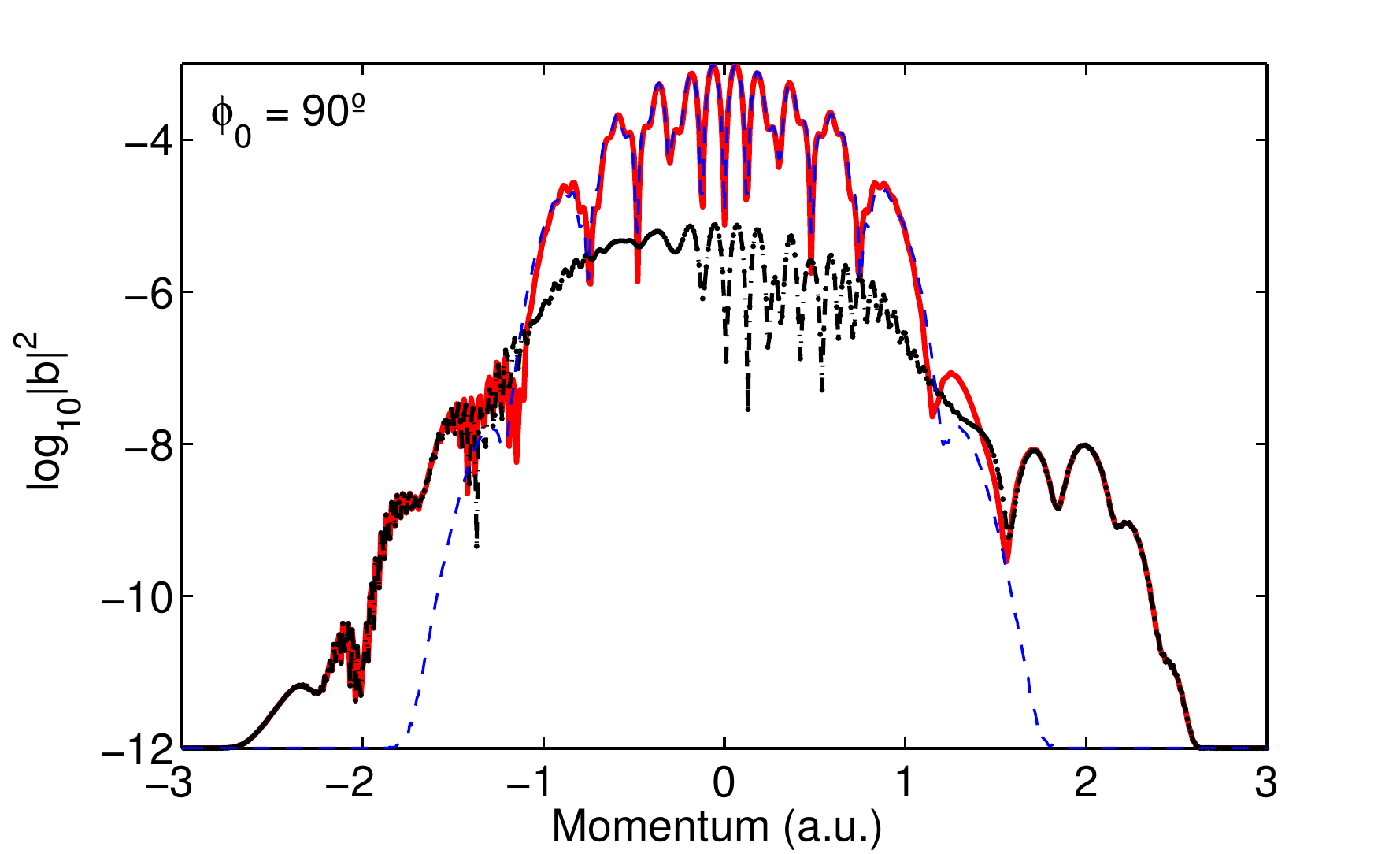}}
    \caption{(color online) Left-right photoelectron momentum distributions (in logarithm scale) for the different contributions, direct term $|b_0(p_z,t_{\rm F})|^2$ (blue dashed line), the re-scattering term $|b_1(p_z,t_{\rm F})|^2$ (black dashed with points line), and the ``total" term, $|b(p_z,t_{\rm F})|^2=|b_0(p_z,t_{\rm F})+b_1(p_z,t_{\rm F})|^2$ (red solid line), to the ATI spectra for two different CEP values, $\phi_0=$ 0\degree and $\phi_0=$ 90\degree  are depicted in (a) and (b), respectively. The laser pulse and the atomic parameters are the same that those used in Fig.~\ref{Fig:1D}.}
    \label{Fig:1DAsymmetric}
\end{figure}

For the energy range $5 U_p \lesssim E_{p_z}\lesssim12 U_p$, the
term $|b_{\rm L}|^2$ is less than $|b_{\rm R}|^2$ around
$\phi_0=0\degree$. This implies that left electron trajectories
have less probability to perform backward re-scattering than those
trajectories emitted to the right. Note that this process changes
if the CEP of the laser pulse is larger than $90\degree$ thereby
the change is for energies between $5 U_p \lesssim
E_{p_z}\lesssim8 U_p$. In this interval the electron trajectories
emitted to the left have a larger probability than the ones
towards. For low-energy photoelectrons $E_{p_z}<5 U_p$, the
asymmetry oscillates between positive and negative values, which
means that the left-right direct photoelectrons are more difficult
to evaluate compared to using re-scattered ones. Thus, these
results depicted in Fig.~\ref{Fig:1DAsimetA} clearly show that our
model describes the typical dependence of the ATI spectra on the
CEP~\cite{NatPaulus2001,Milosevic2006} and in particular the
backward re-scattering events. Our model therefore can be used to
describe the absolute phase of the driving IR laser pulse. With
the purpose to understand the left-right symmetry (or asymmetry)
presented in Fig.~\ref{Fig:1DAsimetA}, we compute both the  direct
and  re-scattering terms  for two  different  CEP  values, namely
$\phi_0= 0\degree$ and $\phi_0= 90\degree$. The results are
depicted in Fig.~\ref{Fig:1DAsymmetric}. In the case of $\phi_0=
0\degree$ the laser pulse is asymmetric with respect to the pulse
envelope maximum, i.e.~it has a $\sin(\omega_0t)$ carrier wave.
Consequently, from the  phase contribution in Eq.~(\ref{Eq:b_0})
of the direct term, one can expect  that  the phase  as a function
of time  is asymmetric as well. It is a consequence of the Fourier
relation that a temporal asymmetry leads to an asymmetric spectral
phase. In analogy, the  temporal asymmetry of the phase of the
direct  photoelectron  term, leads to the final photoelectron
momentum  distribution $|b_0(p_z,t_{\rm F})|^2$ being asymmetric
with respect to the momentum  zero axis.  This dependence is the
origin of the asymmetric shape of the left-right emitted
photoelectrons  shown in Fig.~\ref{Fig:1DAsymmetric}(a). On the
other hand, when the  phase  of Eq.~(\ref{Eq:b_0}) is time
symmetric, which is the case of $\phi_0= 90\degree$, i.e.~the
phase is proportional to $\cos(\omega_0t)$, we infer that the
photoelectron spectrum for the direct term should be symmetric.
This is exactly what we observe in the direct term which is
depicted in Fig.~\ref{Fig:1DAsymmetric}(b). Moreover, in both
cases the re-scattering  term  $|b_1(p_z,t_{\rm F})|^2$ is
asymmetric with respect to the $p_z=0$ momentum. Hence, from  the
quasi-classical analysis addressed in Eq.~(\ref{Eq:B1}) and due to
the few-cycle electric field waveform, the electron trajectories
strongly depend on the CEP and the left-right momentum asymmetry
is visible due to the occurrence and interference of only a few
emission and re-scattering  events.

\begin{figure}[htb]
            \subfigure[$\,$SFA calculations]{\includegraphics[width=0.45\textwidth] {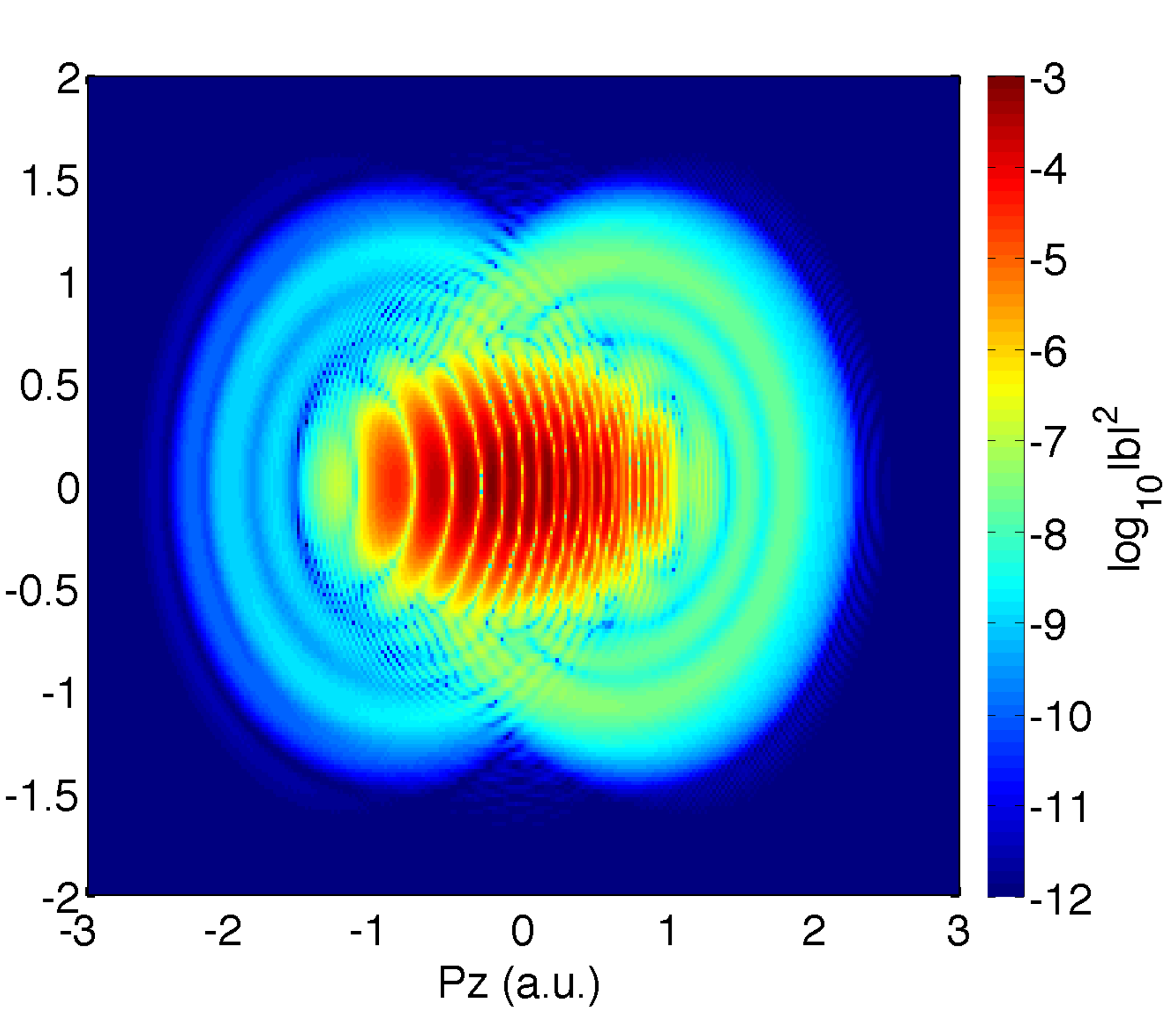}}
     \subfigure[$\,$TDSE calculations]{\includegraphics[width=0.45\textwidth]{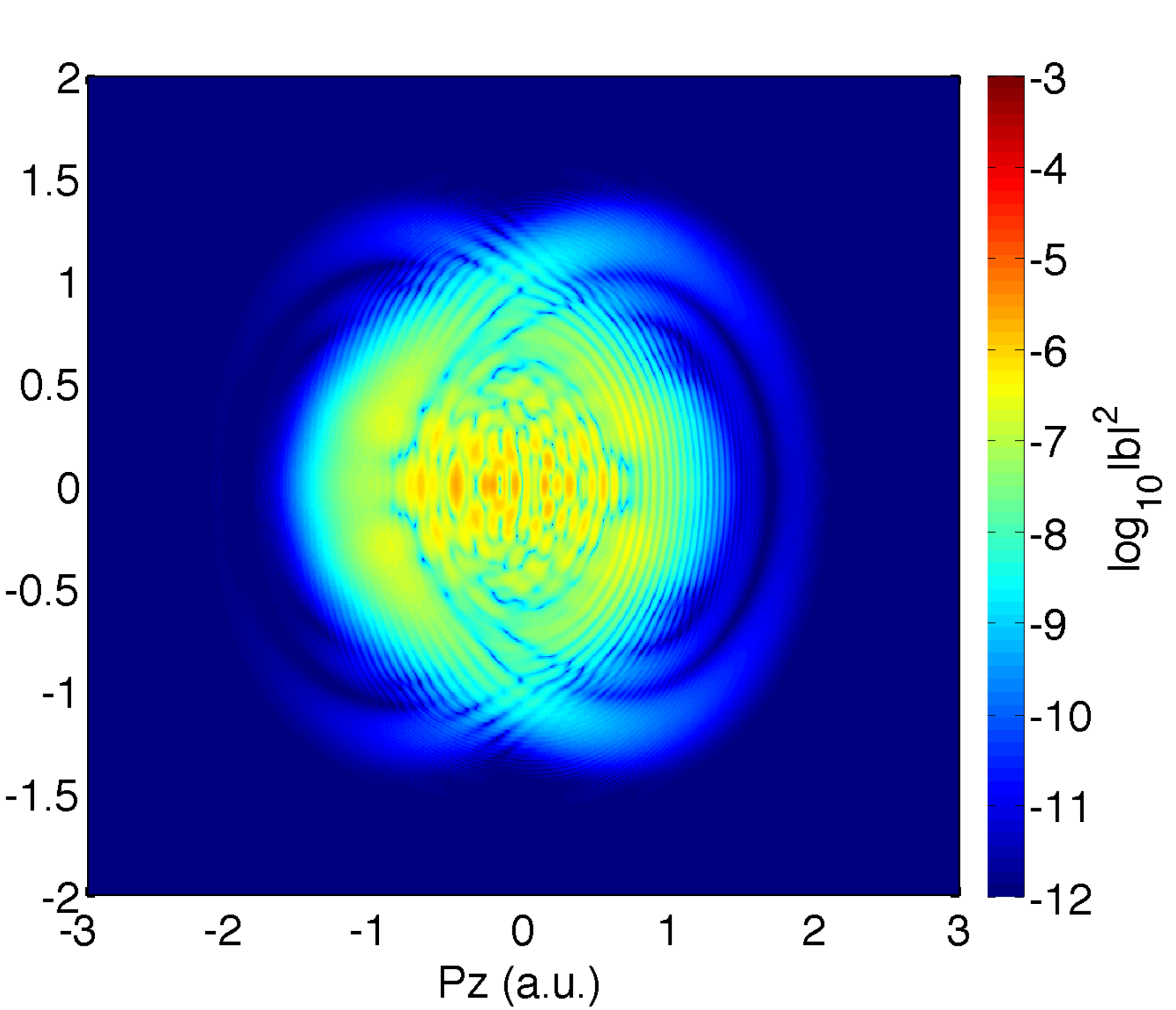}}
    \caption{(color online) Comparison between our semi-classical model and the TDSE in 2D for an hydrogen system. (a)-(b) Photoelectron ATI spectra $|b(p_z,p_y,t_{\rm F})|^2$ (in logarithmic scale) computed by employing our model and
    the TDSE-2D numerical solution, respectively.
    The laser pulse parameters used in these calculations are the same as those employed in Fig~\ref{Fig:1D}.
    Note that the laser field is polarized along the $z$-direction. }
    \label{Fig:ATI2DD}
\end{figure}

In order to complete  the analysis,  we have extended  our
numerical  calculations  of the photoelectron  momentum
distribution for the ATI process from a 1D-momentum line to a
2D-momentum plane. In Fig.~\ref{Fig:ATI2DD} we depict results for
both models: our analytical quasi-classical  ATI model
(Fig.~\ref{Fig:ATI2DD}(a)) and the exact numerical solution of the
TDSE in 2D (Fig.~\ref{Fig:ATI2DD}(b)). We find qualitative good
agreement between the results  of our model and the full numerical
solution of the TDSE in 2D. We also find that the distribution is
symmetric with respect to the $p_y$ axis. Note that these
observations are in good agreement with calculations  and
measurements  presented in
Refs.~\cite{PRLArbo2006,NJPKling2008,NatPhysWittmann2009}.

The comparison shows that our quasi-classical  approach  can  be
used  to  model  2D-momentum distributions and even 3D-momentum
distributions. However, from the contrast between the two models,
we infer that  our  semi-analytical  model is limited to
photoelectrons with high energies. The origin of this discrepancy
arises from the approximation made in the model with regard to the
atomic potential. Statement (iii) relates to the fact that the
atomic potential is neglected when the electron is born in the
continuum. Hence, we expect that electrons with lower final
energies are not well described by our quasi-classical approach.

\begin{figure}[htb]
            \subfigure[~Direct Term]{\includegraphics[width=0.45\textwidth] {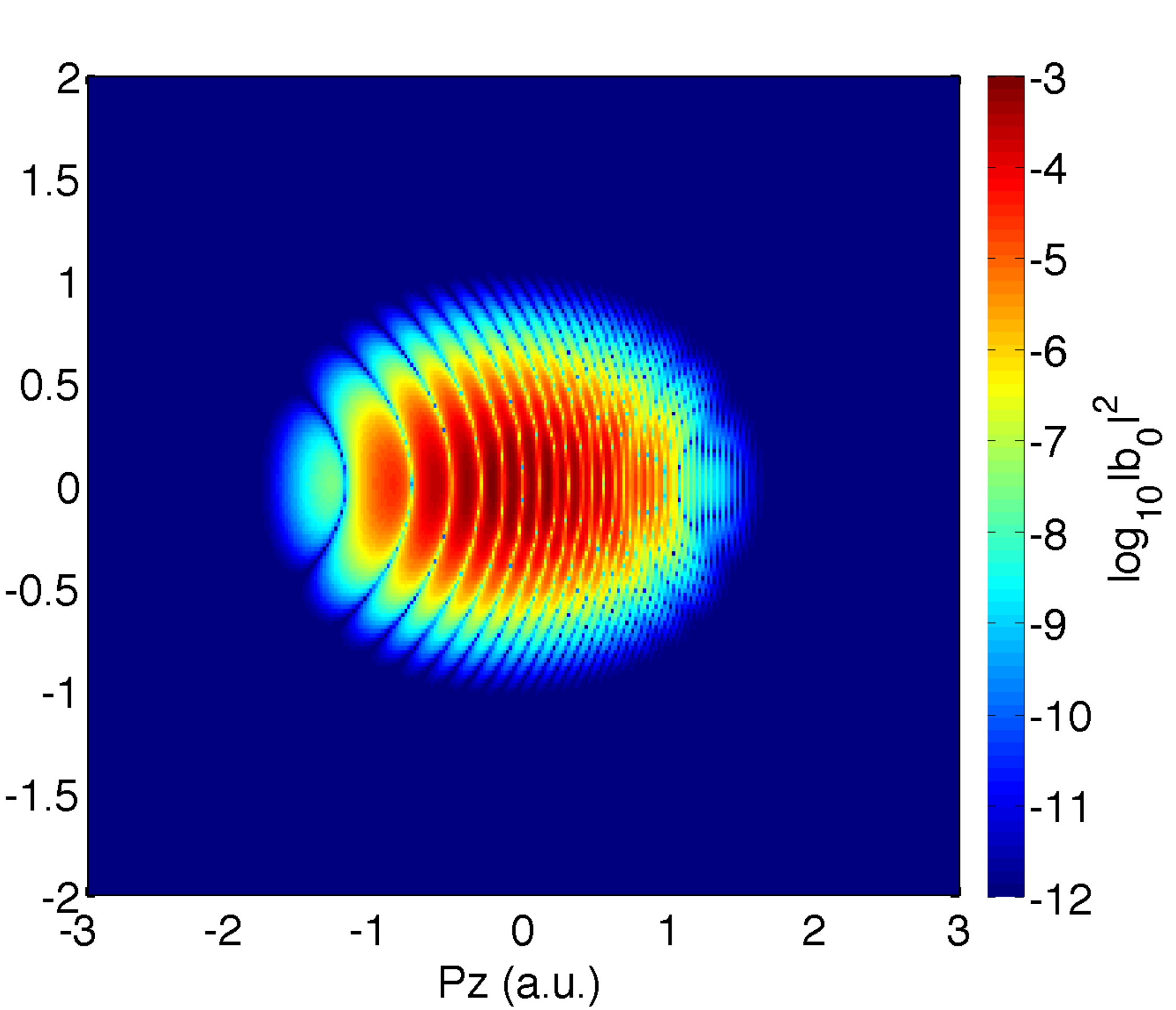}
            \label{fig:subfig}}
     \subfigure[~Rescattering Term]{\includegraphics[width=0.45\textwidth]{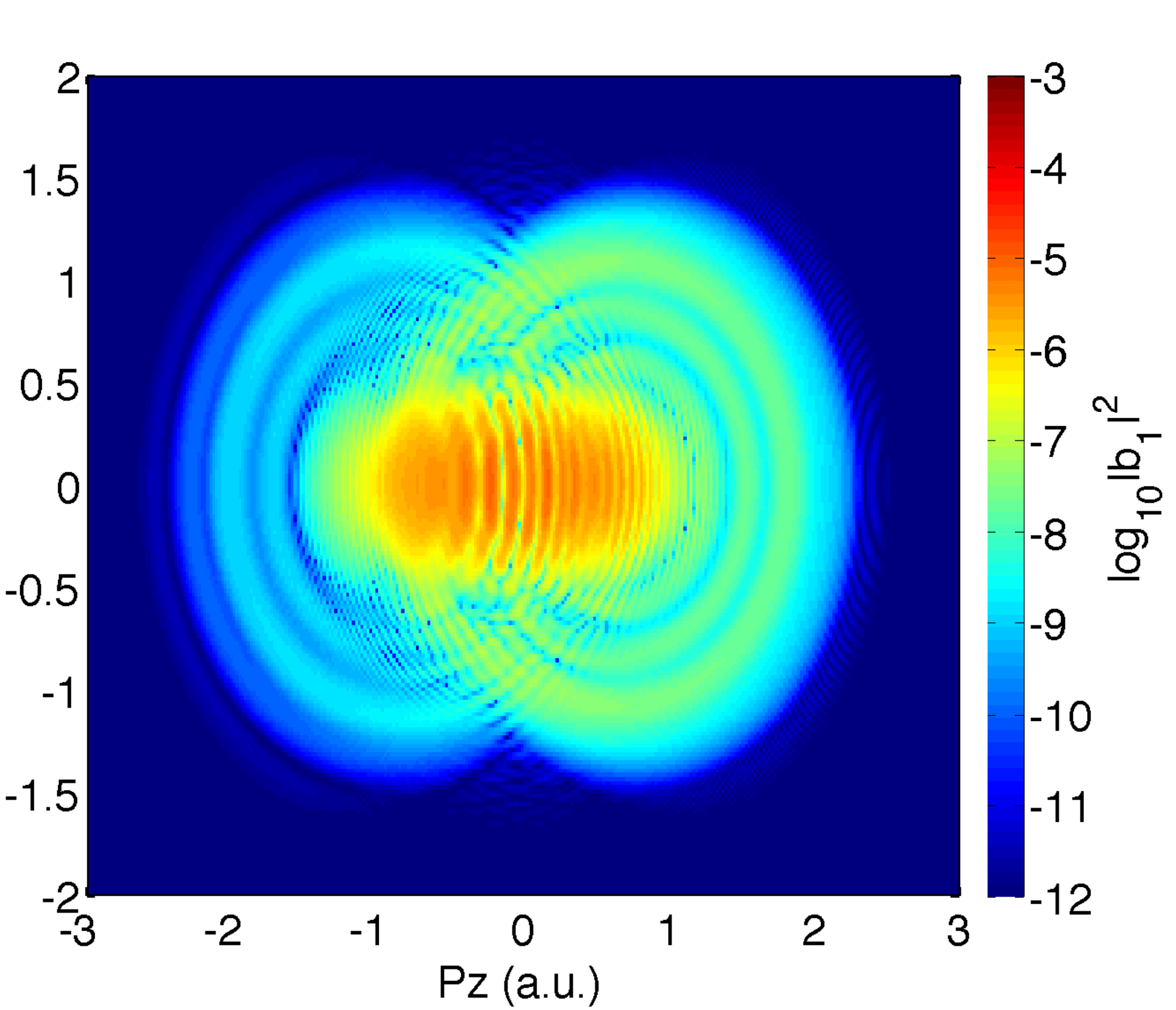}}
     \subfigure[~Interference Term]{\includegraphics[width=0.45\textwidth]{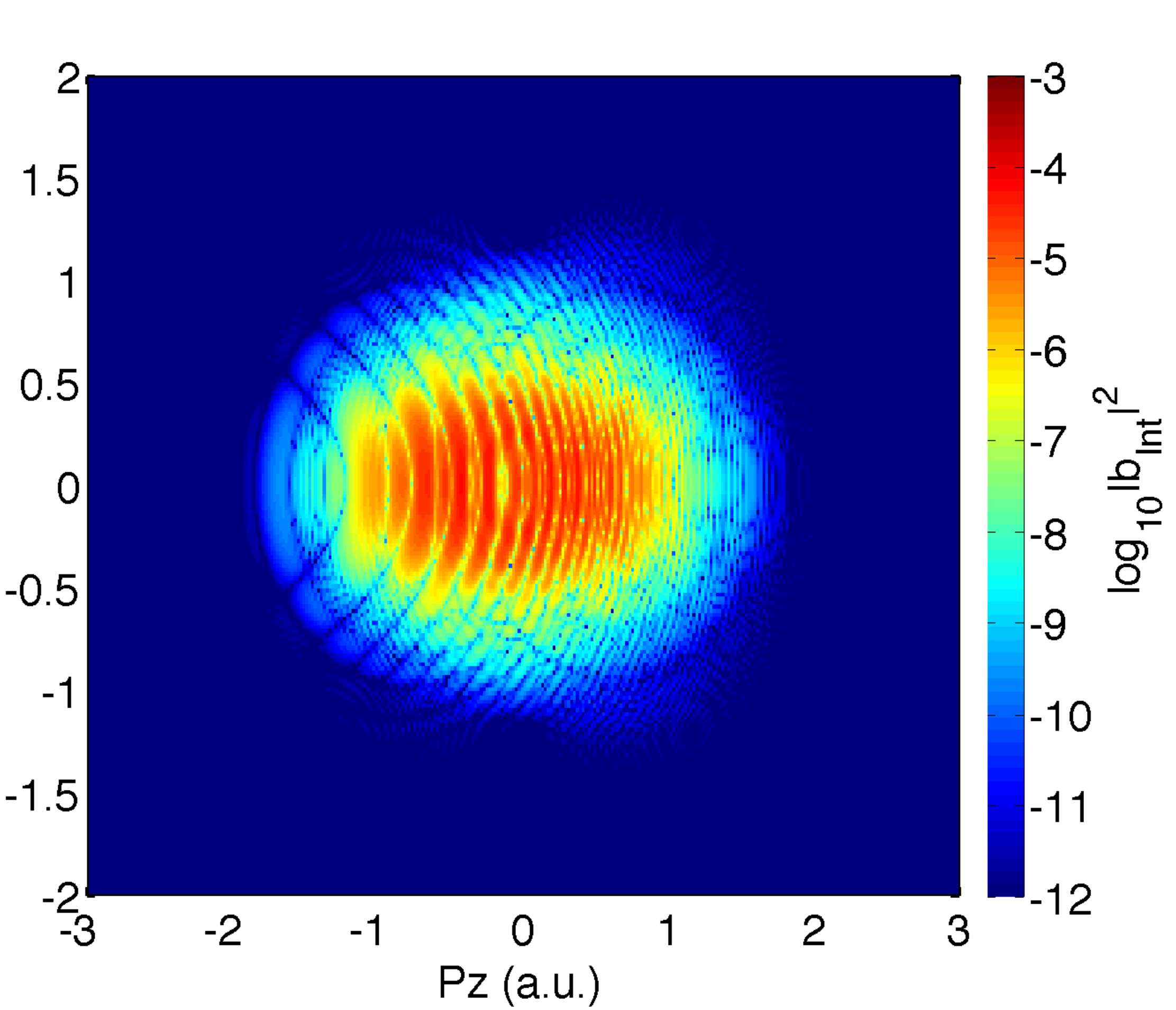}}
     \subfigure[~Total contribution]{\includegraphics[width=0.45\textwidth]{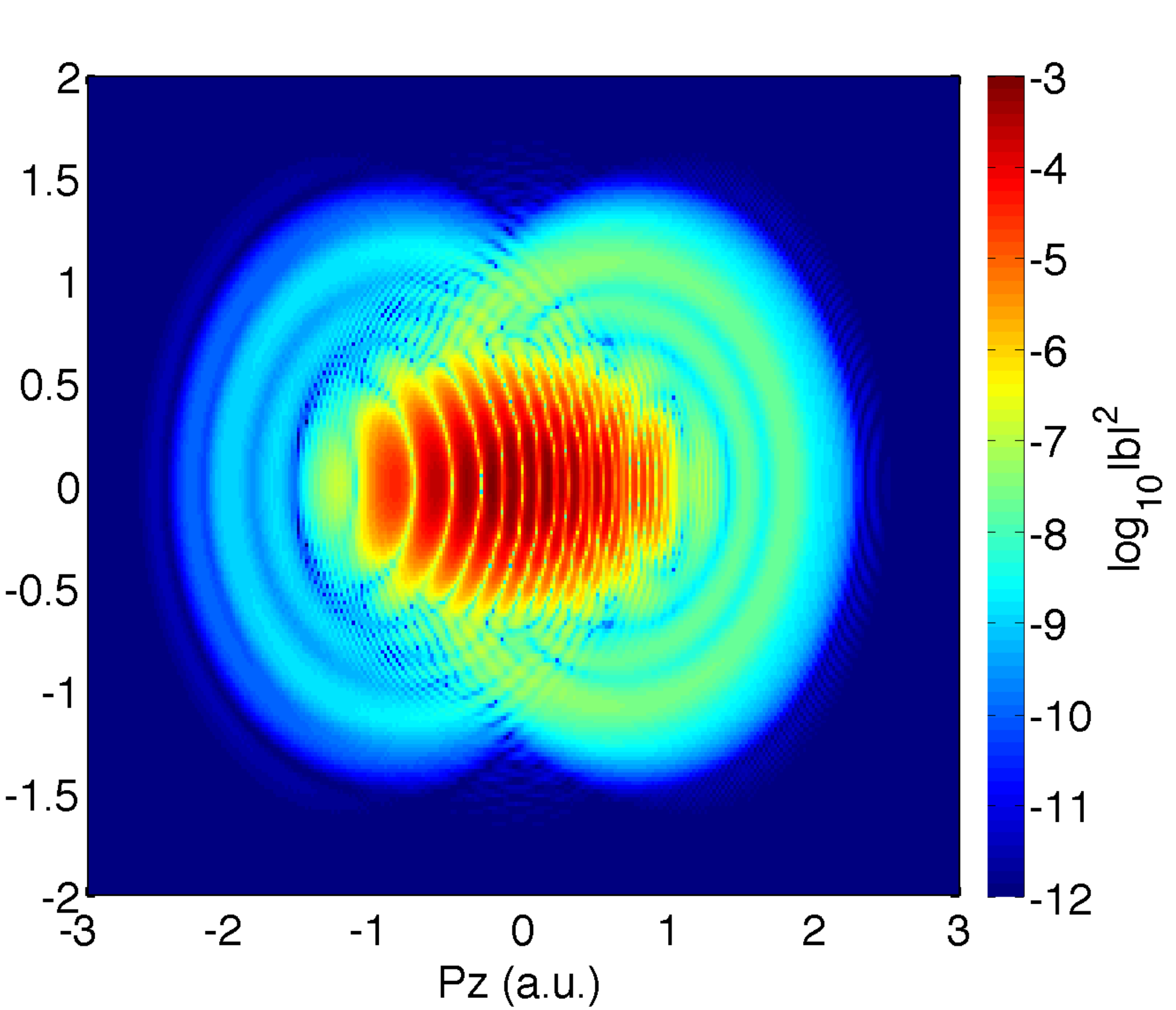}}
    \caption{(color online) Different contributions to the photoelectron spectra for a 2D-momentum plane $(p_z,p_y)$. ATI photoelectron spectra (in logarithmic scale) as a function of the momentum $(p_z,p_y)$ computed by our quasi-classical model for each term. (a) Direct term, (b) re-scattering term, (c) interference term and (d) total contribution.}
    \label{Fig:2DdTerm}
\end{figure}

Finally, the main advantage of the analytical model is shown in
Fig.~\ref{Fig:2DdTerm} which depicts the individual contributions
to the 2D ATI spectrum, namely, the direct
(Fig.~\ref{Fig:2DdTerm}(a)),  re-scattering
(Fig.~\ref{Fig:2DdTerm}(b)) and interference term
(Fig.~\ref{Fig:2DdTerm}(c)), respectively. The resulting total and
experimentally accessible ATI momentum spectrum
$|b({p}_z,p_y,t_{\rm F})|^2$ is shown in Fig.~\ref{Fig:2DdTerm}(d).

The atomic potential and the laser parameters used in these
simulations are identical to those employed in the calculations
for Figs.~\ref{Fig:TermsDRS}-\ref{Fig:1DAsymmetric}. Analog to the
1D calculations, the  computed photoelectron momentum spectrum for
the direct term (Fig.~\ref{Fig:2DdTerm}(a)) shows contributions
for electron energies less than $2U_p$. We find that the
contribution of the  re-scattering term
(Fig.~\ref{Fig:2DdTerm}(b)) extends to higher momentum values.
Clearly visible is the symmetry  of the structures about  the
$p_y$ axis for all the terms and a left-right asymmetric shape for
electrons with $p_z<0$ or $p_z>0$.

We like to emphasize the importance of Eqs.~(\ref{Eq:B1}) and
(\ref{Eq:RSE}), from which we conclude that the form of the
calculated  ATI spectra depends strongly on the parameters
$\Gamma$ and $\gamma$ of our SR potential  model. These parameters
have a strong influence on the re-scattering term, which could get
largely suppressed for a particular choice of them. This strong
dependence suggests that  the re-scattering  process depends
strongly on the atomic target  which means that particular
structural information  is encoded in the ATI photoelectron
spectra. Consequently the proposed  semi-analytical  model can be
used to extract target structure and  electron  dynamics  from
measured  photoelectron  spectra.

\section{Conclusions and Outlook}

We have studied the photoionization process mediated by a strong
laser field interacting with an atomic system.  We analyzed in
detail approximate analytical solutions  of the  TDSE,  obtained
under  the  assumption  of the  strong  field approximation,
i.e.~once an  electron  is tunnel ionized,  its dynamics is solely
governed by the driving  laser, which leads to re-scattering  or
re-collision events. Based on this approach, we have identified
and calculated the  two  main  contributing terms  in the
ionization process: the direct  and  the  re-scattering transition
probability  amplitudes. In addition, the bound-free dipole and
the re-scattering transition matrix elements were {\it
analytically computed} for a non-local potential.
We stressed that this is one of the main difference of our
developed model that those traditionally found in the literature
for the ATI process. These {\it analytical derivations} of the
re-scattering matrix element allowed us to demonstrate that the
re-scattering process strongly depended on the atomic target
features. A quasi-classical analysis of the re-scattering
transition amplitude  was performed in terms  of the  saddle point
approximation, which permits linking the dynamics to relevant
quasi-classical  information, i.e.~classical electron
trajectories.  Our analytical results suggested that  the main
contributions to the re-scattering transition amplitude correspond
to electron trajectories,  with significant probability  of
backward  scattering  off the ionic core.

Our model was used to demonstrate that both contributions, the
direct  and the re-scattering terms, shown left-right asymmetry
depending on the carrier envelope phase of the laser pulse. This
behavior has been confirmed by a comparison with the exact
numerical  solution  of the TDSE and we found very qualitative
good agreement, particularly in the high energy region of the
photoelectron  spectra.  Apart from testing the validity of our
model, we stress that it presents important advantages, such as
the possibility to disentangle  the effects of both  the direct
and re-scattered terms.

We showed also that the model is sensitive to the CEP,  and by
using the fact that we can investigate individual contributions to
the photoelectron spectrum, we identified the re-scattering term
that  plays a dominant role by varying its influence based on the
atomic parameters. These findings confirm that the photoelectron
spectra contain structural information  about  the re-scattering
process, i.e.~about  the shape of the ground state wave function,
as well as the  bound-free,  and  free-free matrix  transition
elements. This dependence implies that atomic  structural
information  can be efficiently extracted with our model for
methods such as LIED which measure ATI spectra.

While our aim was the establishment of a basic semi-analytical
theoretical framework based on the SFA, we note that our approach
is applicable to more complex, and thus more interesting  systems
such as molecules. The method could be extended to describe
dynamically  evolving molecular systems or atomic clusters. It
will be interesting to corroborate which kind of information could
be extracted, and whether one could visualize molecular dynamics
such as vibrations or dissociation, or if the ground state
molecular orbital could be reconstructed. We will address these
and similar questions in future publications.

\acknowledgments{This work was supported by Ministerio de
Econom\'{\i}a y Competitividad through Plan Nacional
(FIS2011-30465-C02-01) and FrOntiers of QUantum Sciences (FOQUS):
Atoms, Molecules, Photons and Quantum Information
(FIS2013-46768-P), the Catalan Agencia de Gestio d'Ajuts
Universitaris i de Recerca (AGAUR) with SGR 2014-2016, Fundaci\'o
Cellex Barcelona and funding from LASERLAB-EUROPE, Grant agreement
284464. N.S. was supported by the Erasmus Mundus Doctorate Program
Europhotonics (Grant No. 159224-1-2009-1-FR-ERA MUNDUS-EMJD). A.
C. and M. L. acknowledge ERC AdG OSYRIS. J.B. acknowledges
FIS2014-51478-ERC.

%

\end{document}